\documentclass[10pt,revtex]{article}
\usepackage{rotating}
\usepackage{epsfig,amssymb,amsmath}

\input{epsf}

\def\laq{\raise 0.4 ex\hbox{$<$}\kern -0.8 em\lower 0.62 ex\hbox{$\sim$}}
\def\gaq{\raise 0.4 ex\hbox{$>$}\kern -0.7 em\lower 0.62 ex\hbox{$\sim$}}

\def\beq{\begin{equation}}
\def\eeq{\end{equation}}
\def\beqa{\begin{eqnarray}}
\def\eeqa{\end{eqnarray}}

\def\to{\rightarrow}

\def\ev{\mbox{eV}}

\def\tev{\mbox{TeV}}

\def\mpc{\mbox{Mpc}}

\def\to{\rightarrow}

\def\ev{\mbox{eV}}

\def\tev{\mbox{TeV}}

\def\mpc{\mbox{Mpc}}

\def\AJ{{\it Ap. J.} }
\def\AJL{{\it Ap. J. Lett.} }

\def\AJS{{\it Ap. J. Supp.} }

\def\AP{{\it Ann. Phys.} }

\def\CQG{{\it Class. Quantum Gravity} }

\def\GRG{{\it Gen. Relativity and Gravit.} }

\def\IJMP{{\it Int. J. Mod. Phys.} }

\def\JHEP{{\it JHEP} }
\def\JP{{\it J. Phys.} }

\def\MPL{{\it Mod. Phys. Lett.} }
\def\MNRAS{{\it Mon. Not. R. Ast. Soc.} }
\def\NAT{{\it Nature} }

\def\NC{{\it Il Nuovo Cimento} }
\def\NJP{{\it New. Journal Phys.} }
\def\NP{{\it Nucl. Phys.} }
\def\PL{{\it Phys. Lett.} }
\def\PMAG{{\it Philos. Mag.} }
\def\PR{{\it Phys. Rev.} }
\def\PRL{{\it Phys. Rev. Lett.} }
\def\PRTS{{\it Physics Reports} }

\def\PTP{{\it Progr. Theor. Phys.} }

\def\RMP{{\it Rev. Mod. Phys.} }

\def\RPP{{\it Rep. Prog. Phys.} }

\def\be{\beta}
\def\ga{\gamma}

\def\th{\theta}

\def\ka{\kappa}

\def\De{\Delta}

\def\La{\Lambda}

\def\Om{\Omega}

\def\vev#1{\langle {#1}\rangle}

 \def\frac#1#2{{\textstyle{{#1}\over {#2}}}}
 \def\lsim{\mathrel{\rlap{\lower4pt\hbox{\hskip1pt$\sim$}}
    \raise1pt\hbox{$<$}}} \def\gsim{\mathrel{\rlap{\lower4pt\hbox{\hskip1pt$\sim$}}
    \raise1pt\hbox{$>$}}}
\def\sqr#1#2{{\vcenter{\vbox{\hrule height.#2pt
         \hbox{\vrule width.#2pt height#1pt \kern#1pt
         \vrule width.#2pt}
         \hrule height.#2pt}}}}

\def\gappeq{\mathrel{\rlap {\raise.5ex\hbox{$>$}} {\lower.5ex\hbox{$\sim$}}}}

\def\lappeq{\mathrel{\rlap{\raise.5ex\hbox{$<$}}
{\lower.5ex\hbox{$\sim$}}}}

\begin{document}
\pagestyle{plain}

\begin{center}

{\Large\bf General Theory of Relativity: Will it survive the next decade?}

\vspace*{1.0cm}

{Orfeu Bertolami,$^a$  Jorge P\'aramos,$^a$ and Slava G. Turyshev$^b$}\\
\vspace*{0.5cm}
{\it $^a$Instituto Superior T\'ecnico, Departamento de F\'\i sica, \\
Av. Rovisco Pais, 1049-001 Lisboa, Portugal}\\
{\it $^b$Jet Propulsion Laboratory,
California Institute of Technology, \\
4800 Oak Grove Drive, Pasadena, CA 91109, USA}
\end{center}

\vspace*{0.9cm}

\begin{abstract}
The nature of gravity is fundamental to our understanding of our own solar system, the galaxy and the structure and evolution of the Universe. Einstein's general theory of relativity is the standard model that is used for almost ninety years to describe gravitational phenomena on these various scales.  We review the foundations of general relativity, discuss the recent progress in the tests of relativistic gravity, and present motivations for high-accuracy gravitational experiments in space.  We also summarize the science objectives and technology needs for the laboratory experiments in space with laboratory being the entire solar system. We discuss the advances in our understanding of fundamental physics anticipated in the near future and evaluate discovery potential for the recently proposed gravitational experiments.

\end{abstract}

%********************* CONTENTS
%{\small \tableofcontents}
%\listoffigures
%\listoftables
%********************* Section 1 **********************
\section{Introduction}
\label{sec:intro}

To understand the Universe in its vast and complex splendor seems
a daunting task, yet human curiosity and wonder over centuries and
civilizations have always led humankind to seek answers to some of
the most compelling questions of all -- How did the Universe come
to be? What is it made of? What forces rule its behavior? Why is
it the way it is? What will ultimately become of it? With its
prominent influence on natural phenomena at every distance scale,
gravitation plays a pivotal role in this intellectual quest.

Gravity was known to humans long before the present-day picture of
four fundamental interactions was formed. The nature of gravity is
fundamental to our understanding of our solar system, the galaxy
and the structure and evolution of the Universe. It was Newton who
first understood that gravity not only dictates the fall of apples
and all bodies on Earth, but also planetary motion in our solar
system and the sun itself are govern by the same physical
principles. On the larger scales the effects of gravity are even
more pronounced, guiding the evolution of the galaxies, galactic
clusters and ultimately determining the fate of the Universe.
Presently the Einstein's general theory of relativity is a key to
the understanding a wide range of phenomena, spanning from the
dynamics of compact astrophysical objects such as neutron stars
and black holes, to cosmology, where the Universe itself is the
object of study. Its striking predictions include gravitational
lensing and waves, and only black holes have not yet been directly
confirmed.

The significance that general relativity (GR) plays for our
understanding of nature, makes the theory a focus of series of
experimental efforts performed with ever increasing accuracy.
However, even after more than ninety years since general
relativity was born, Einstein's theory has survived every test.
Such longevity does not mean that it is absolutely correct, but
serves to motivate more precise tests to determine the level of
accuracy at which it is violated. This motivates various precision
tests of gravity both in laboratories and in space; as a result,
we have witnessed an impressive progress in this area over the
last two decades. However, there are a number of reasons to
question the validity of this theory, both theoretical and
experimental.

On the theoretical front, the problems arise from several
directions, most dealing with the strong gravitational field
regime; this includes the appearance of spacetime singularities and the
inability to describe the physics of very strong gravitational
fields using the standard of classical description. A way out of
this difficulty would be attained through gravity quantization. However, despite
the success of modern gauge field theories in describing the
electromagnetic, weak, and strong interactions, it is still not
understood how gravity should be described at the quantum level.
Our two foundational theories of nature, quantum mechanics and
GR, are not compatible with each other. In
theories that attempt to include gravity, new long-range forces
can arise in addition to the Newtonian inverse-square law. Even at
the classical level, and assuming the Equivalence Principle,
Einstein's theory does not provide the most general way to
establish the spacetime metric. Regardless of whether the
cosmological constant should be included, there are also important
reasons to consider additional fields, especially scalar fields.
Although the latter naturally appear in these modern theories,
their inclusion predicts a non-Einsteinian behavior of gravitating
systems. These deviations from GR lead to a violation of the Equivalence Principle, a foundation of general relativity, modification of large-scale gravitational phenomena, and cast doubt upon the constancy of the fundamental ``constants.'' These predictions motivate new searches for very small deviations of relativistic gravity from GR and provide a new theoretical paradigm and guidance for further gravity experiments.

Meanwhile, on the experimental front, recent cosmological
observations has forced us to accept the fact that our current
understanding of the origin and evolution of the Universe is at
best incomplete, and possibly wrong. It turned out that, to our
surprise, most of the energy content of the Universe resides in
presently unknown dark matter and dark energy that may permeate
much, if not all of spacetime. If so, then this dark matter may be
accessible to laboratory experimentation. It is likely that the
underlying physics that resolve the discord between quantum
mechanics and GR will also shed light on
cosmological questions addressing the origin and ultimate destiny
of the Universe. Recent progress in the development of vastly
superior measurement technology placed fundamental physics in a
unique position to successfully address these vital questions.
Moreover, because of the ever increasing practical significance of
the general theory of relativity (i.e. its use in spacecraft
navigation, time transfer, clock synchronization, etalons of time,
weight and length, etc.) this fundamental theory must be tested to
increasing accuracy.

This paper is organized as follows: Section~\ref{sec:gr-survey}
discusses the foundations of the general theory of relativity and 
reviews the results of the recent experiments designed to test the 
foundations of this theory. Section~\ref{sec:beyond} presents 
motivations for extending the theoretical model of gravity provided 
by GR; it presents a model arising from string theory, 
discusses the scalar-tensor theories of gravity, and also highlights 
phenomenological implications of these proposals. This section 
also reviews the motivations and the search for new interactions of 
nature and discusses the hypothesis of gravitational shielding. 
Section~\ref{cosmological} addresses the astrophysical and 
cosmological phenomena that led to some recent proposals that modify gravity 
on large scales; it discusses some of these proposals and reviews 
their experimental implications. Section \ref{sec:space} discusses 
future missions and experiments aiming to expand our knowledge of 
gravity. Finally, conclusions and an outlook are presented.

%%%%%%%%%%%%%%%%%%%%%%%%%% Section 2 %%%%%%%%%%%%%%%%%%%%%%%%%%%%%
\section{Testing Foundations of General Relativity}
\label{sec:gr-survey}

General relativity began its empirical success in 1915, by
explaining the anomalous perihelion precession of Mercury's orbit,
using no adjustable theoretical parameters. Shortly thereafter,
Eddington's 1919 observations of stellar lines-of-sight during a
solar eclipse confirmed the doubling of the deflection angles
predicted by the Einstein's theory, as compared to Newtonian-like
and Equivalence Principle arguments; this made the theory an
instant success. From these beginnings, GR
has been extensively tested in the solar system, successfully
accounting for all data gathered to date. Thus, microwave ranging
to the Viking Lander on Mars yielded accuracy $\sim 0.2$ in the
tests of GR \cite{viking_shapiro1,viking_reasen,viking_shapiro2}.
Spacecraft and planetary radar observations reached an accuracy of
$\sim 0.15$ \cite{anderson02}. The astrometric observations of
quasars on the solar background performed with Very-Long Baseline
Interferometry improved the accuracy of the tests of gravity to
$\sim 0.045$
\cite{RoberstonCarter91,Lebach95,Shapiro_SS_etal_2004}. Lunar
laser ranging $\sim 0.011 $ verification of GR via
precision measurements of the lunar orbit
\cite{Ken_LLR68,Ken_LLR91,Ken_LLR30years99,Ken_LLR_PPNprobe03,JimSkipJean96,Williams_etal_2001,Williams_Turyshev_Boggs_2004}. Finally, the recent 
experiments with the Cassini spacecraft
improved the accuracy of the tests to $\sim 0.0023 $
\cite{cassini_ber}. As a result, GR became the
standard theory of gravity when astrometry and spacecraft
navigation are concerned.

To date, GR is also in agreement with the data
collected from the binary millisecond pulsars. In fact, recently a
considerable interest has been shown in the physical processes
occurring in the strong gravitational field regime with
relativistic pulsars providing a promising possibility to test
gravity in this qualitatively different dynamical environment. The
general theoretical framework for pulsar tests of strong-field
gravity was introduced in \cite{DamourTaylor92}; the observational
data for the initial tests were obtained with PSR1534
\cite{Taylor_etal92}. An analysis of strong-field gravitational
tests and their theoretical justification was presented in
\cite{Damour_EFarese96a,Damour_EFarese96b,Damour_EFarese98}. The
recent analysis of the pulsar data tested GR to
$\sim 0.04 $ at a $3 \sigma$ confidence level
\cite{lange_etal2001}.

In this Section we present the framework used to plan and analyze
the data in a weak-field and slow motion approximation which is
appropriate to describe dynamical conditions in the solar system.

\subsection{Metric Theories of Gravity and PPN Formalism}

Within the accuracy of modern experiments, the weak-field and slow
motion approximation provides a useful starting point for testing
the predictions of different metric theories of gravity in the
solar system. Following Fock \cite{Fock1,Fock2} and Chandrasekhar
\cite{Chandrasekhar_65}, a matter distribution in this
approximation is often represented by the perfect fluid model with
the density of energy-momentum tensor $\widehat{T}^{mn}$ as given
below:
%{}
\beq \widehat{T}^{mn}=\sqrt{-g}\Big(\Big[\rho_{0}( 1 + \Pi) +
p\Big]u^{m} u^{n} - p g^{mn} \Big), \label{eq:perfect-fl} \eeq
%{}
\noindent where $\rho_0$ is the mass density of the ideal fluid in
coordinates of the co-moving frame of reference, $u^k = {d z^k / d
s}$ are the components of invariant four-velocity of a fluid
element, and $p(\rho)$ is the isentropic pressure connected with
$\rho$ by an equation of state. The quantity $\rho\Pi$ is the
density of internal energy of an ideal fluid. The definition of
$\Pi$ results from the first law of thermodynamics, through the
equation $ u^n\big(\Pi_{;n} + p\big({1/ {\widehat
\rho}}\big)_{;n}\big) = 0 $, where the subscript $;n$ denotes a
covariant derivative, ${\widehat \rho}=\sqrt{-g}\rho_0u^0$ is the
conserved mass density (see further details in Refs.
\cite{Fock2,Chandrasekhar_65,Brumberg,Will1}). Given the
energy-momentum tensor, one finds the solutions of the
gravitational field equations for a particular theory of
gravity.\footnote{A powerful approach developing a weak-field
approximation for GR was presented in Refs.
\cite{DSX,Blanchet_etal_95,Damour_Vokrouhlicky_95}. It combines an
elegant ``Maxwell-like'' treatise of the spacetime metric in both
the {\it global} and {\it local} reference frames with the
Blanchet-Damour multipole formalism \cite{Blanchet_Damour_86}.
This approach is applicable for an arbitrary energy-stress tensor
and is suitable for addressing problems of strong field regime. Application of this method to a general N-body problem in a weak-field and slow motion approximation was developed in Ref.~\cite{Kopeikin_Vlasov_2004}.}

Metric theories of gravity have a special position among all the
other possible theoretical models. The reason is that,
independently of the many different principles at their
foundations, the gravitational field in these theories affects the
matter directly through the metric tensor $g_{mn}$, which is
determined from the field equations. As a result, in contrast to
Newtonian gravity, this tensor expresses the properties of a
particular gravitational theory and carries information about the
gravitational field of the bodies.

Generalizing on a phenomenological parameterization of the
gravitational metric tensor field, which Eddington originally
developed for a special case, a method called the parameterized
post-Newtonian (PPN) metric has been developed
\cite{Nordtvedt_1968a,Will_1971,Will_Nordtvedt_1972}. This method
represents the gravity tensor's potentials for slowly moving
bodies and weak inter-body gravity, and is valid for a broad class
of metric theories, including GR as a unique case.
The several parameters in the PPN metric expansion vary from
theory to theory, and they are individually associated with
various symmetries and invariance properties of the underlying
theory. Gravity experiments can be analyzed in terms of the PPN
metric, and an ensemble of experiments will determine the unique
value for these parameters, and hence the metric field itself.

As we know it today, observationally, GR is the most successful
theory so far as solar system experiments are concerned (see e.g.
\cite{Will2005} for an updated review). The implications of GR for
solar system gravitational phenomena are best addressed via the
PPN formalism for which the metric tensor of the general
Riemannian spacetime is generated by some given distribution of matter in the form of an ideal fluid, given by  Eq.~(\ref{eq:perfect-fl}). It is represented by a sum of gravitational potentials with arbitrary coefficients, the PPN parameters. If, for simplicity, one assumes that Lorentz invariance, local position invariance and total momentum conservation hold, the metric tensor in four dimensions in the so-called PPN-gauge may be written\footnote{Note the geometrical units $\hbar=c=G=1$ are used throughout, as is the metric signature convention $(-+++)$.} as {}
\beqa g_{00}&=&-1+2U - 2\beta\, U^2 + 2(\gamma+1)\Phi_1
+2\Big[(3\gamma+1-2\beta)\Phi_2+\Phi_3+3\gamma\Phi_4\Big]+ {\cal
O}(c^{-5}), \\ \nonumber g_{0i} & = &-{1\over 2}(4\gamma+3)V_i-
{1\over2}W_i + {\cal O}(c^{-5}), \qquad
g_{ij}=\delta_{ij}(1+2\gamma U)+ {\cal O}(c^{-4}).
\label{eqno(1)} \eeqa

The order of magnitude of the various terms is
determined according to the rules $U \sim v^2 \sim \Pi \sim p/\rho
\sim \epsilon$, $v^i \sim |d/dt|/|d/dx| \sim \epsilon^{1/2}$. The
parameter $\gamma$ represents the measure of the curvature of the
spacetime created by the unit rest mass; the parameter $\beta$ is
the measure of the non-linearity of the law of superposition of
the gravitational fields in a theory of gravity or the measure of
the metricity. The generalized gravitational potentials,
proportional to $ U^2 $, result from integrating the energy-stress
density, Eq.~(\ref{eq:perfect-fl}), are given by {}
%{}
\beq U({\bf x},t) = \int d^3{\bf x}'{\rho_0 ({\bf x}',t) \over
|{\bf x}-{\bf x}'|}, \qquad V^\alpha({\bf x},t) = - \int d^3{\bf
x}'{\rho_0({\bf x}',t)v^\alpha ({\bf x}',t)\over |{\bf x}-{\bf
x}'|},\eeq
%{}
\beq W^i({\bf x},t) = \int d^3{\bf x}'\rho_0({\bf x}',t)v_j({\bf
x}',t) {(x^j-x'^j)(x^i-x'^i)\over|{\bf x}-{\bf x}'|^3}, \eeq
%{}
\beq \Phi_1({\bf x},t) = - \int d^3{\bf x}'{\rho_0({\bf x}',t) v^2
({\bf x}',t)\over |{\bf x}-{\bf x}'|},\qquad \Phi_{2} ({\bf x}',t)
= \int d^3{\bf x}' {\rho_0({\bf x}',t)U({\bf x}',t)\over|{\bf
x}-{\bf x}'|}, \eeq
%{}
\beq \Phi_{3}({\bf x},t) = \int d^3{\bf x}'{ \rho_0({\bf x}',t)
\Pi({\bf x}',t)\over |{\bf x}-{\bf x}'|}d^3z'^\nu, \qquad
\Phi_{4}({\bf x},t) = \int d^3{\bf x}' {p({\bf x}',t)\over |{\bf
x}-{\bf x}'|}. \label{eq:gen-pots} \eeq

In the complete PPN framework, a particular metric theory of
gravity in the PPN formalism with a specific coordinate gauge
might be fully characterized by means of ten PPN parameters
\cite{Will1,Turyshev96}. Thus, besides the parameters $\gamma,
\beta$, there other eight parameters $\alpha_1, \alpha_2,
\alpha_3, \zeta, \zeta_1,\zeta_2,\zeta_3,\zeta_4$. The formalism
uniquely prescribes the values of these parameters for each
particular theory under study. In the standard PPN gauge
\cite{Will1} these parameters have clear physical meaning, each
quantifying a particular symmetry, conservation law or fundamental
tenant of the structure of spacetime. Thus, in addition to the
parameters $\gamma$ and $\beta$ discussed above, the group of
parameters $\alpha_1, \alpha_2, \alpha_3$ specify the violation of
Lorentz invariance (or the presence of the privileged reference
frame), the parameter $\zeta$ quantifies the violation of the
local position invariance, and, finally, the parameters
$\zeta_1,\zeta_2,\zeta_3,\zeta_4$ reflect the violation of the law
of total momentum conservation for a closed gravitating system.
Note that GR, when analyzed in standard PPN gauge,
gives: $\gamma=\beta=1$ and all the other eight parameters vanish.
The Brans-Dicke theory \cite{Brans} is the best known of the
alternative theories of gravity. It contains, besides the metric
tensor, a scalar field and an arbitrary coupling constant
$\omega$, which yields the two PPN parameter values, $\beta=1$,
$\gamma= ( 1 + \omega ) / ( 2 + \omega )$, where $\omega$ is an unknown
dimensionless parameter of this theory. More general scalar tensor
theories (see Section \ref{sec:vacuum}) yield values of $\beta$
different from one
\cite{Damour_Nordtvedt_1993a}.

The main properties of the PPN metric tensor given by
Eqs.~(\ref{eqno(1)})-(\ref{eq:gen-pots}) are well established and
widely in use in modern astronomical practice
\cite{Moyer71,Moyer81,Turyshev96,Brumberg,Standish_etal_92,Will1}.
For practical purposes one uses this metric to generate the
equations of motion for the bodies of interest. These equations
are then used to produce numerical codes in relativistic orbit
determination formalisms for planets and satellites
\cite{Moyer81,Standish_etal_92,Turyshev96} as well as for
analyzing the gravitational experiments in the solar system
\cite{Will1,turyshev_acfc_2003}.

In what follows, we discuss the foundations of general theory of
relativity together with our current empirical knowledge on their
validity. We take the standard approach to GR
according to which the theory is supported by the following basic
tenants:

\begin{enumerate}

\item[1).]  Equivalence Principle (EP), which states that 
freely falling bodies do have the same acceleration 
in the same gravitational field independent on their compositions, 
which is also known as 
the principle of universality of the 
free fall (discussed in Section~\ref{sec:eep});

\item[2).] Local Lorentz invariance (LLI), which suggests that 
clock rates are independent on the 
clock's velocities (discussed in Section~\ref{LLI});

\item[3).] Local position invariance (LPI), which 
postulates that clocks rates are also independent 
on their spacetime positions (discussed in Section~\ref{LPI}).

\end{enumerate}

\subsection{The Equivalence Principle (EP)}
\label{sec:eep}

Since Newton, the question about the equality of inertial and
passive gravitational masses has risen in almost every theory of
gravitation. Thus, almost one hundred years ago Einstein
postulated that not only mechanical laws of motion, but also all
non-gravitational laws should behave in freely falling frames as
if gravity was absent. It is this principle that predicts
identical accelerations of compositionally different objects in
the same gravitational field, and also allows gravity to be viewed
as a geometrical property of spacetime--leading to the general
relativistic interpretation of gravitation.

Below we shall discuss two different ``flavors'' of the
Equivalence Principle, the weak and the strong forms of the EP
that are currently tested in various experiments performed with
laboratory tests masses and with bodies of astronomical sizes.

\subsubsection{The Weak  Equivalence Principle (WEP)}
\label{sec:wep}

The weak form of the EP (the WEP) states that the gravitational
properties of strong and electro-weak interactions obey the EP. In
this case the relevant test-body differences are their fractional
nuclear-binding differences, their neutron-to-proton ratios, their
atomic charges, \textit{etc}.. Furthermore, the equality of gravitational
and inertial masses implies that different neutral massive test
bodies will have the same free fall acceleration in an external
gravitational field, and therefore in freely falling inertial
frames the external gravitational field appears only in the form
of a tidal interaction \cite{Singe_1960}. Apart from these tidal
corrections, freely falling bodies behave as if external gravity
is absent \cite{Anderson_etal_1996}.

According to GR, the light rays propagating near a gravitating
body are achromatically scattered by the curvature of the
spacetime generated by the body's gravity field. The entire
trajectory of the light ray is bent towards the body by an angle
depending on the strength of the body's gravity. In the solar
system, the sun's gravity field produces the largest effect,
deflecting the light by as much as $1.75'' \cdot (R_{\odot} / b)$, 
where $R_{\odot}$ is the solar radius and $b$ is the
impact parameter. The Eddington's 1919 experiment confirmed the fact that 
photons obey the laws of free fall in a gravitational field as 
predicted by GR. The original accuracy was only 10\% which
was recently improved to 0.0023\% by a solar conjunction
experiment performed with the Cassini spacecraft
\cite{cassini_ber}. 

The Pound-Rebka experiment, performed in 1960,
further verified effects of gravity on light by testing the
universality of gravity-induced frequency shift, $\De \nu$, that follows from
the WEP:
\beq {\Delta \nu \over \nu} = {g h \over c^2} = (2.57 \pm 0.26)
\times 10^{-15}, \label{eq:1.15} \eeq \noindent
\noindent where $g$ is the acceleration of gravity and $h$ the height of
fall \cite{Pound-Rebka}.

The WEP can be scrutinized by studying the free fall of
antiprotons and antihydrogen, even though the experimental
obstacles are considerable; the subject has been extensively
reviewed in Ref. \cite{Nieto}. This would help investigating to
what extent does gravity respect the fundamental CPT symmetry of
local quantum field theories, namely if antiparticles fall as
particles in a gravitational field. As we shall see later, CPT
symmetry may be spontaneously broken in some string/M-theory
vacua's; some implications of this will also be mentioned in the
context of the validity Local Lorentz invariance. The ATHENA
(ApparaTus for High precision Experiments on Neutral Antimatter)
and the ATRAP collaborations at CERN have developed techniques to deal with the
difficulties of storing antiprotons and creating an antihydrogen
atom (see Refs. \cite{Athena,Atrap} for recent accounts), but no 
gravitational has been performed so far. 
On the other hand, the former CPLEAR Collaboration has reported on
a test of the WEP involving neutral kaons \cite{Apostolakis}, with
limits of $6.5$, $4.3$ and $1.8 \times 10^{-9}$ respectively for
scalar, vector and tensor potentials originating from the sun with
a range much greater than 1~AU acting on kaons and antikaons.
Despite their relevance, these results say nothing about new
forces that couple to the baryon number, and therefore are at best
complementary to further tests yet to be performed with
antiprotons and antihydrogen atoms.

Most extensions to GR are metric in nature, that is, they assume that
the WEP is valid. However, as emphasized by
\cite{Damour_1996,Damour_2001}, almost all extensions to the
standard model of particle physics generically predict new forces
that would show up as apparent violations of the EP; this occurs
specially in theories containing macroscopic-range quantum fields
and thus predicting quantum exchange forces that generically
violate the WEP, as they couple to generalized ``charges'', rather
than to mass/energy as does gravity
\cite{Damour_Polyakov_1994a}.

In a laboratory, precise tests of the EP can be made by comparing
the free fall accelerations, $a_1$ and $a_2$, of different test
bodies. When the bodies are at the same distance from the source
of the gravity, the expression for the equivalence principle takes
the elegant form
\beq {\Delta a \over a} = {2(a_1- a_2) \over a_1 + a_2} =
\left({M_G \over M_I}\right)_{\hskip -3pt 1} -\left({M_G \over
M_I}\right)_{\hskip -3pt 2} = \Delta\left({M_G \over
M_I}\right), \label{WEP_da} \eeq
\noindent where $M_G$ and $M_I$ are the gravitational and inertial
masses of each body. The sensitivity of the EP test is determined
by the precision of the differential acceleration measurement
divided by the degree to which the test bodies differ (\textit{e.g.}
composition).

The WEP has been subject to various laboratory tests throughout
the years. In 1975, Collela, Overhauser and Werner \cite{Collela}
showed with their interferometric experiment that a neutron beam
split by a silicon crystal traveling through distinct
gravitational paths interferes as predicted by the laws of quantum
mechanics, with a gravitational potential given by Newtonian
gravity, thus enabling an impressive verification of the WEP
applied to an elementary hadron. Present-day technology has achieved 
impressive limits for the interferometry of atoms rising against 
gravity, of order $3 \times 10^{-8}$ \cite{Kasevich}.

Various experiments have been performed to measure the ratios of
gravitational to inertial masses of bodies. Recent experiments on
bodies of laboratory dimensions verify the WEP to a fractional
precision~$\Delta(M_G/M_I) \lesssim 10^{-11}$ by
\cite{Roll_etal_1964}, to $\lesssim 10^{-12}$ by
\cite{Braginsky_Panov_1972,Su_etal_1994} and more recently to a
precision of $\lesssim 1.4\times 10^{-13}$ \cite{Adelberger_2001}.
The accuracy of these experiments is sufficiently high to confirm
that the strong, weak, and electromagnetic interactions each
contribute equally to the passive gravitational and inertial
masses of the laboratory bodies. A review of the most recent
laboratory tests of gravity can be found in Ref. \cite{Gundlach}.

Quite recently, Nesvizhevsky and collaborators have reported
evidence for the existence of gravitational bound states of
neutrons \cite{Nesvizhevsky}; the experiment was, at least conceptually, 
put forward long ago, in 1978 \cite{Luschikov}. Subsequent steps
towards the final experiment are described in Ref.
\cite{Nesvizhevsky1}. This consists in allowing ultracold neutrons
from a source at the Institute Laue-Langevin reactor in Grenoble
to fall towards a horizontal mirror under the influence of the
Earth's gravitational field. This potential confines the motion of
the neutrons, which do not move continuously vertically, but
rather jump from one height to another as predicted by quantum
mechanics. It is reported that the minimum measurable energy is of
$1.4 \times 10^{-12}~\ev$, corresponding to a vertical velocity of
$1.7$~cm/s. A more intense beam and an enclosure mirrored on all
sides could lead to an energy resolution down to $10^{-18}~\ev$.

We remark that this experiment opens fascinating perspectives,
both for testing non-commutative versions of quantum mechanics, as
well as the connection of this theory with gravity
\cite{Bertolami18}. It also enables a new criteria for the
understanding the conditions for distinguishing quantum from
classical behavior in function of the size of an observed system
\cite{Bertolami19}.

This impressive evidence of the WEP for laboratory bodies is incomplete for
astronomical body scales. The experiments searching for WEP
violations are conducted in laboratory environments that utilize
test masses with negligible amounts of gravitational self-energy
and therefore a large scale experiment is needed to test the
postulated equality of gravitational self-energy contributions to
the inertial and passive gravitational masses of the bodies
\cite{Nordtvedt_1968a}. Recent analysis of the lunar laser ranging
data demonstrated that no composition-dependent acceleration
effects \cite{Baessler_etal_1999} are present.

Once the self-gravity of the test bodies is non-negligible
(currently with bodies of astronomical sizes only), the
corresponding experiment will be testing the ultimate version of
the EP -- the strong equivalence principle, that is discussed
below.

\subsubsection{The Strong Equivalence Principle (SEP)}
\label{sec:sep}

In its strong form the EP is extended to cover the gravitational
properties resulting from gravitational energy itself. In other
words, it is an assumption about the way that gravity begets
gravity, i.e. about the non-linear property of gravitation.
Although GR assumes that the SEP is exact,
alternate metric theories of gravity such as those involving
scalar fields, and other extensions of gravity theory, typically
violate the SEP
\cite{Nordtvedt_1968a,Nordtvedt_1968b,Ken_LLR68,Nordtvedt_1991}.
For the SEP case, the relevant test body differences are the
fractional contributions to their masses by gravitational
self-energy. Because of the extreme weakness of gravity, SEP test
bodies that differ significantly must have astronomical sizes.
Currently, the Earth-Moon-Sun system provides the best solar
system arena for testing the SEP.

A wide class of metric theories of gravity are described by the
parameterized post-Newtonian formalism
\cite{Nordtvedt_1968b,Will_1971,Will_Nordtvedt_1972}, which allows
one to describe within a common framework the motion of celestial
bodies in external gravitational fields. Over the last 35 years,
the PPN formalism has become a useful framework for testing the
SEP for extended bodies. To facilitate investigation of a possible
violation of the SEP, in that formalism the ratio between
gravitational and inertial masses, $M_G/M_I$, is expressed
\cite{Nordtvedt_1968a,Nordtvedt_1968b} as
\beq \left[{M_G \over M_I}\right]_{\tt SEP} = 1 +
\eta\left({\Omega \over Mc^2}\right), \label{eq:MgMi} \eeq
\noindent where $M$ is the mass of a body, $\Omega $ is the body's
(negative) gravitational self-energy, $Mc^2$ is its total
mass-energy, and $\eta$ is a dimensionless constant for SEP
violation \cite{Nordtvedt_1968a,Nordtvedt_1968b,Ken_LLR68}.
Any SEP violation is quantified by the parameter $\eta$: in
fully-conservative, Lorentz-invariant theories of gravity
\cite{Will1,Will_2001} the SEP parameter is related to the PPN
parameters by $ \eta = 4\beta - \gamma -3$. In GR
$\gamma = \beta = 1$, so that $\eta = 0$.

The self energy of a body $B$ is given by
\beq \left({\Omega  \over Mc^2}\right)_B = - {G \over 2 M_B
c^2}\int_B d^3{\bf x} d^3{\bf y} {\rho_B({\bf x})\rho_B({\bf y})
\over | {\bf x} - {\bf y}|}. \label{eq:omega} \eeq
\noindent For a sphere with a radius $R$ and uniform density,
$\Omega /Mc^2 = -3GM/5Rc^2 = -3 v_E^2/10 c^2$, where $v_E$ is the
escape velocity. Accurate evaluation for solar system bodies
requires numerical integration of the expression of
Eq.~(\ref{eq:omega}). Evaluating the standard solar model
\cite{Ulrich_1982} results in $(\Omega /Mc^2)_\odot \sim -3.52 \times
10^{-6}$. Because gravitational self-energy is proportional to
$M^2$ and also because of the extreme weakness of gravity, the
typical values for the ratio $(\Omega /Mc^2)$ are $\sim 10^{-25}$
for bodies of laboratory sizes. Therefore, the experimental
accuracy of a part in $10^{13}$ \cite{Adelberger_2001} which is so
useful for the WEP is not sufficient to test on how gravitational
self-energy contributes to the inertial and gravitational masses
of small bodies. To test the SEP one must consider planetary-sized
extended bodies, where the ratio Eq.~(\ref{eq:omega}) is
considerably higher.

Nordtvedt \cite{Nordtvedt_1968a,Ken_LLR68,Nordtvedt_1970}
suggested several solar system experiments for testing the SEP.
One of these was the lunar test. Another, a search for the SEP
effect in the motion of the Trojan asteroids, was carried out by
\cite{Orellana_Vucetich_1988,Orellana_Vucetich_1993}.
Interplanetary spacecraft tests have been considered by
\cite{Anderson_etal_1996} and discussed
\cite{Anderson_Williams_2001}. An experiment employing existing
binary pulsar data has been proposed \cite{Damour_Schafer_1991}.
It was pointed out that binary pulsars may provide an excellent
possibility for testing the SEP in the new regime of strong
self-gravity
\cite{Damour_EFarese96a,Damour_EFarese96b},
however the corresponding tests have yet to reach competitive
accuracy \cite{Wex_2001,Lorimer_Freire_2004}.

To date, the Earth-Moon-Sun system has provided the most accurate
test of the SEP; recent analysis of LLR data test the EP to a high
precision, yielding $\Delta (M_G/M_I)_{\tt EP}
=(-1.0\pm1.4)\times10^{-13}$ \cite{Williams_Turyshev_Boggs_2004}.
This result corresponds to a test of the SEP of $\Delta
(M_G/M_I)_{\tt SEP} =(-2.0\pm2.0)\times10^{-13}$ with the SEP
violation parameter $\eta=4\beta-\gamma-3$ found to be
$\eta=(4.4\pm4.5)\times 10^{-4}$. Using the recent Cassini result
for the PPN parameter $\gamma$, PPN parameter $\beta$ is
determined at the level of $\beta-1=(1.2\pm1.1)\times 10^{-4}$ 
(see more details in \cite{Williams_Turyshev_Boggs_2004}).

\subsection{Local Lorentz Invariance (LLI)}
\label{LLI}

Invariance under Lorentz transformations states that the laws of
physics are independent of the frame velocity; this is an underlying 
symmetry of all current
physical theories. However, some evidence recently found in the
context of string field theory indicates that this symmetry can be
spontaneously broken. Naturally, the experimental verification of
this breaking poses a significant challenge. It has already been
pointed out that astrophysical observations of distant sources of gamma
radiation could hint what is the nature of gravity-induced wave
dispersion in vacuum \cite{Amelino1, Biller} and therefore points 
towards physics beyond the Standard Model of Particles
and Fields (hereafter -- Standard Model). Limits on Lorentz
symmetry violation based on the observations of high-energy cosmic
rays with energies beyond $5 \times 10^{19}~\ev$, the so-called
Greisen-Zatsepin-Kuzmin (GKZ) cut-off \cite{Greisen}, have also
been discussed \cite{Sato,Coleman,Mestres,Bertolami2}.

A putative violation of Lorentz symmetry has been a repeated
object of interest in the literature. A physical description of
the effect of our velocity with respect to a presumably preferred
frame of reference relies on a constant background cosmological
vector field, as suggested in \cite{Phillips}. Based on the
behavior of the renormalization group $\beta$-function of
non-abelian gauge theories, it has also been argued that Lorentz
invariance could be just a low-energy symmetry \cite{Nielsen}.

Lorentz symmetry breaking due to non-trivial solutions of string
field theory was first discussed in Ref. \cite{Kostelecky1}. These
arise from the string field theory of open strings and may have
implications for low-energy physics. For instance, assuming that
the contribution of Lorentz-violating interactions to the vacuum
energy is about half of the critical density implies that feeble
tensor-mediated interactions in the range of $ \sim 10^{-4}$~m should
exist \cite{Bertolami3}. Furthermore, Lorentz violation may induce
the breaking of conformal symmetry; this, together with inflation
may explain the origin of the primordial magnetic fields required
to explain the observed galactic magnetic field \cite{Bertolami4}.
Also, violations of Lorentz invariance may imply in a breaking of
the fundamental CPT symmetry of local quantum field theories
\cite{Kostelecky}. Quite remarkably, this can be experimentally
verified in neutral-meson \cite{Kostelecky2}
experiments,\footnote{These CPT violating effects are unrelated
with those due to possible non-linearities in quantum mechanics,
presumably arising from quantum gravity and already investigated
by the CPLEAR Collaboration \cite{Adler}.} Penning-trap
measurements \cite{Bluhm1} and hydrogen-antihydrogen spectroscopy
\cite{Bluhm2}. This spontaneous breaking of CPT symmetry allows
for an explanation of the baryon asymmetry of the Universe: in the
early Universe, , after the breaking of the Lorentz and CPT
symmetries, tensor-fermion interactions in the low-energy limit of
string field theories give rise to a chemical potential that
creates in equilibrium a baryon-antibaryon asymmetry in the
presence of baryon number violating interactions
\cite{Bertolami5}.

Limits on the violation of Lorentz symmetry are available from
laser interferometric versions of the Michelson-Morley experiment,
by comparing the velocity of light, $c$ and the maximum attainable
velocity of massive particles, $c_i$, up to $\delta \equiv
|c^2/c_{i}^2 - 1| < 10^{-9}$ \cite{Brillet}. More accurate tests
can be performed via the Hughes-Drever experiment \cite{Hughes,
Drever}, where one searches for a time dependence of the
quadrupole splitting of nuclear Zeeman levels along Earth's orbit.
This technique achieves an impressive limit of $\delta < 3 \times
10^{-22}$ \cite{Lamoreaux}. A recent reassessment of these results
reveals that more stringent bounds can be reached, up to 8 orders
of magnitude higher \cite{Kostelecky3}. The parameterized
post-Newtonian parameter $\alpha_{3}$ can be used to set
astrophysical limits on the violation of momentum conservation and
the existence of a preferred reference frame. This parameter,
which vanishes identically in GR can be accurately determined from
the pulse period of pulsars and millisecond pulsars
\cite{Will2005, Bell}. The most recent results yields a limit 
on the PPN parameter $\alpha_3$ 
of $|\alpha_{3}| < 2.2 \times 10^{-20}$ \cite{BellD}.

After the cosmic microwave background radiation (CMBR) has been
discovered, an analysis of the interaction between the most
energetic cosmic-ray particles and the microwave photons was
mandatory. As it turns out, the propagation of the
ultra-high-energy nucleons is limited by inelastic collisions with
photons of the CMBR, preventing nucleons
with energies above $5 \times 10^{19}~\ev$ from reaching Earth
from further than 50--100~$\mpc$. This is the already mentioned
GZK cut-off \cite{Greisen}. However, events where the estimated
energy of the cosmic primaries is beyond the GZK cut-off where
observed by different collaborations
\cite{Yoshida,Bird,Brooke,Efimov}. It has been suggested
\cite{Sato,Coleman} (see also \cite{Mestres}) that slight
violations of Lorentz invariance would cause energy-dependent
effects that would suppress otherwise inevitable processes such as
the resonant scattering reaction, $p + \ga_{2.73K} \to
\De_{1232}$. The study of the kinematics of this process produces
a quite stringent constraint on the validity of Lorentz invariance, 
$\delta < 1.7 \times 10^{-25}$ \cite{Bertolami2,Bertolami6}.

Quite recently, the High Resolution Fly's Eye collaboration
suggested that the gathered data show that the GZK cutoff holds
for their span of observations \cite{Abbasi}. Confirmation of this
result is of great importance, and the coming into operation of
the Auger collaboration \cite{Auger} in the near future will
undoubtedly provide further insight into this fundamental
question. It is also worth mentioning that the breaking of Lorentz
invariance can occur in the context of non-commutative field
theories \cite{Carroll}, even though this symmetry may hold (at
least) at first non-trivial order in perturbation theory of the
non-commutative parameter \cite{Bertolami15}. Actually, the idea
that the non-commutative parameter may be a Lorentz tensor has
been considered in some field theory models \cite{Bertolami16}.
Also, Lorentz symmetry can be broken in loop quantum gravity
\cite{Gambini}, quantum gravity inspired spacetime foam scenarios
\cite{Garay} or via the spacetime variation of fundamental
coupling constants \cite{Lehnert}. For a fairly complete review
about Lorentz violation at high-energies the reader is directed to
Ref.~\cite{Mattingly}. Note that a gravity model where LLI is
spontaneously broken was proposed in Ref.
\cite{Kostelecky4,Kostelecky5} and solutions where discussed in
Ref. \cite{Bertolami-Paramos2005a}.

\subsubsection{Spontaneous Violation of Lorentz Invariance}

The impact of a spontaneous violation of Lorentz invariance on
theories of gravity is of great interest. In this context, the
breaking of Lorentz invariance may be implemented, for instance,
by allowing a vector field to roll to its vacuum expectation
value. This mechanism requires that the potential that rules the
dynamics of the vector field possesses a minimum, in the way
similar to the Higgs mechanism \cite{Kostelecky4}. This,
so-called, ``bumblebee'' vector thus acquires an explicit
(four-dimensional) orientation, and Lorentz symmetry is
spontaneously broken. The action of the bumblebee model is written
as

\beq S = \int d^4 x \sqrt{-g} \left[{1 \over 2 \ka} \left( R + \xi
B^\mu B^\nu R_{\mu\nu} \right) - {1 \over 4} B^{\mu\nu} B_{\mu\nu}
- V(B^\mu B_\mu \pm b^2 ) \right]~~, \eeq

\noindent where $B_{\mu\nu} = \partial_\mu B_\nu -
\partial_\nu B_\mu$, $\xi$ and $b^2$ are a real coupling constant and a free
real positive constant, respectively. The potential $V$ driving
Lorentz and/or CPT violation is supposed to have a minimum at $
B^\mu B_\mu \pm b^2 = 0 $, $V'(b_\mu b^\mu)=0$. Since one assumes
that the bumblebee field $B_\mu$ is frozen at its vacuum
expectation value, the particular form of the potential driving
its dynamics is irrelevant. The scale of $b_\mu$, which controls
the symmetry breaking, must be derived from a fundamental theory,
such as string theory or from a low-energy extension to the
Standard Model; hence one expects either $b$ of order of the Planck mass, $M_P = 1.2 \times 10^{19}$~GeV, or of order of the electroweak
breaking scale, $M_{EW} \simeq 10^2$~GeV.

Previously, efforts to quantify an hypothetical breaking of
Lorentz invariance were primarily directed towards the
phenomenology of such spontaneous Lorentz symmetry breaking in
particle physics. Only recently its implications for gravity have
been more thoroughly explored \cite{Kostelecky4, Kostelecky5}. In
that work, the framework for the LSB gravity model is set up,
developing the action and using the \textit{vielbein} formalism. A
later study \cite{Bertolami-Paramos2005a} considered consequences
of such a scenario, assuming three plausible cases: i) the
bumblebee field acquires a purely radial vacuum expectation value,
ii) a mixed radial and temporal vacuum expectation value and iii)
a mixed axial and temporal vacuum expectation value.

In the first case, an exact black hole solution was found,
exhibiting a deviation from the inverse square law such that the
gravitational potential of a point mass at rest depends on the
radial coordinate as $r^{-1+p}$ where $p$ is a parameter related
to the fundamental physics underlying the breaking of Lorentz
invariance. This solution has a removable singularity at a horizon
of radius $r_s = (2M r_0^{-p})^{1/(1-p)}$, slightly perturbed with
respect to the usual Scharzschild radius $r_{s0} = 2M$, which
protects a real singularity at $r=0$. Known deviations from
Kepler's law yield $p \leq 2 \times 10^{-9}$.

In the second case, no exact solutions was discovered, although a
perturbative method allowed for the characterization of the
Lorentz symmetry breaking in terms of the PPN parameters $ \be
\approx 1 - (K + K_r)/ M$ and $\gamma \approx 1 - (K + 2 K_r) /
M$, directly proportional to the strength of the induced effect,
given by $K$ and $K_r \sim K$, where $K$ and $K_r$ are integration
constants arising from the perturbative treatment of the timelike
spontaneous LSB superimposed on the vacuum Scharzschild metric. An
analogy with Rosen's bimetric theory yields the parameter $\gamma
\simeq (A + B)d$, where $d$ is the distance to the central body
and $A$ and $B$ rule the temporal and radial components of the
vector field vacuum expectation value.

In the third case, a temporal/axial vacuum expectation value for
the bumblebee vector field clearly breaks isotropy, thus
forbidding a standard PPN analysis. However, for the case of the
breaking of Lorentz invariance occurring in the $x_1$ direction,
similar direction-dependent PPN-like parameters were defined as $
\gamma_1 \simeq b^2 \cos^2 \th /2$ and $\gamma_2 \simeq a^2 b^2
\cos^2 \th / 4$, where $a$ and $b$ are proportional respectively
to the temporal and axial components of the vacuum expectation
value acquired by the bumblebee. This enables a crude estimative
of the measurable PPN parameter $\gamma$, yielding $ \gamma
\approx b^2(1-e^2)/4$, where $e$ is the orbit's eccentricity. A
comparison with experiments concerning the anisotropy of inertia
yields $ |b| \leq 2.4 \times 10^{-11} $ \cite{Lamoreaux}.

\subsection{Local Position Invariance (LPI)}
\label{LPI}

Given that both the WEP and LLI postulates have been tested with
great accuracy, experiments concerning the universality of the
gravitational red-shift measure the level to which the LPI holds.
Therefore, violations of the LPI would imply that the rate of a
free falling clock would be different when compared with a
standard one, for instance on the Earth's surface. The accuracy to
which the LPI holds as an invariance of Nature can be
parameterized through $ \Delta \nu / \nu = (1 + \mu) U / c^2$. 
From the already mentioned Pound-Rebka experiment (cf. Eq.
(\ref{eq:1.15})) one can infer that $\mu \simeq 10^{-2}$; the most
accurate verification of the LPI was performed by Vessot and
collaborators, by comparing hydrogen-maser frequencies on Earth
and on a rocket flying to altitude of $10^4$~km  \cite{Vessot}.
The resulting bound is $ \vert \mu \vert < 2 \times 10^{-4}$.
Recently, an one order of magnitude improvement was attained, thus
establishing the most stringent result on the LPI so far
\cite{Bauch}, $ \vert \mu \vert < 2.1 \times 10^{-5}$.

\subsection{Summary of Solar System Tests of Relativistic Gravity}
\label{tests}

\begin{table}[t!]
\caption{The accuracy of determination of the PPN parameters 
$\gamma$ and $\beta$ \cite{Williams_Turyshev_Boggs_2004,Will2005,turyshev_acfc_2003}.
\vspace*{0.5cm}}

\label{ppntable}

\begin{center}
\begin{tabular}{|c|c|c|}
\hline
  PPN parameter & Experiment & Result \\ \hline\hline
  $\ga -1 $ & Cassini 2003 spacecraft radio-tracking &
$ 2.3 \times 10^{-5}$ \\\cline{2-3}
  ~  & Observations of quasars with Astrometric VLBI & $ 3 \times 10^{-4} $ \\\hline 
  $ \be - 1 $ & Helioseismology bound on perihelion shift & $ 3 \times 10^{-3} $  \\\cline{2-3}
  ~ & LLR test of the SEP, assumed: $\eta = 4 \be - \ga - 3$   & $ 1.1 \times 10^{-4} $  \\
  ~ &  and the Cassini result for PPN $\gamma$    &  ~ \\
\hline

\end{tabular}
\\[10pt]
\end{center}

\end{table}

Although, these available experimental data fit quite well with
GR, while allowing for the existence of putative
extensions of this successful theory, provided any new effects are
small at the post-Newtonian scale \cite{Will1}.We shall here
discuss the available phenomenological constraints for alternative
theories of gravity.

\begin{figure}[t]
\centering
\leavevmode\epsfxsize=11.5cm \epsfbox{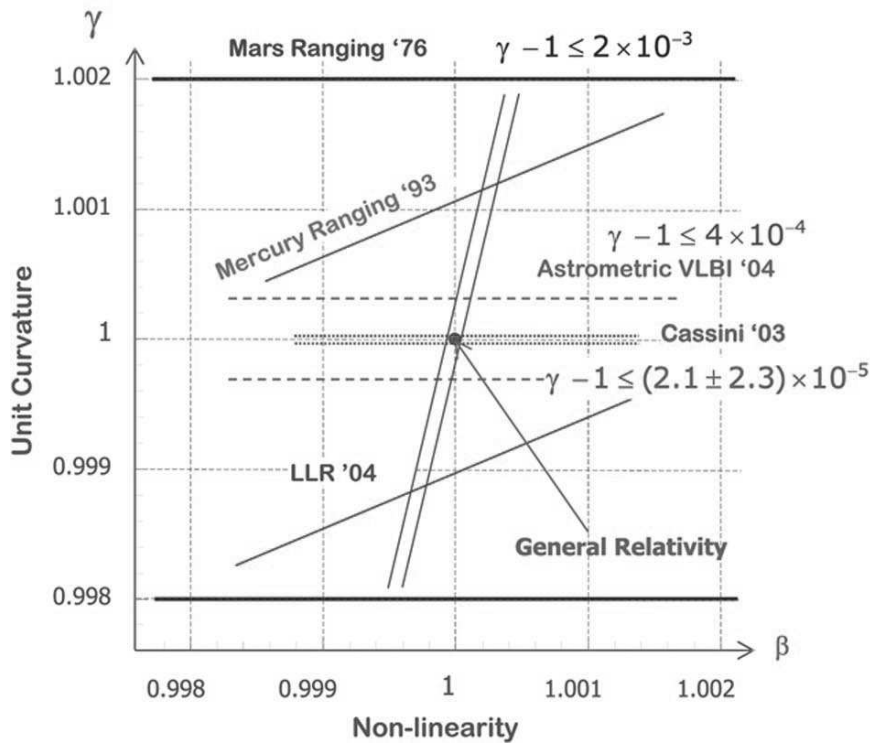}\\
\caption{\label{fig:ppn} The progress in determining the PPN
parameters $\gamma$ and $\beta$ for the last 30 years (adopted
from \cite{Laser_Clocks_LATOR}).}
\end{figure}

Lunar Laser Ranging (LLR) investigates the SEP by looking for a
displacement of the lunar orbit along the direction to the sun.
The equivalence principle can be split into two parts: the WEP
tests the sensitivity to composition and the SEP checks the
dependence on mass. There are laboratory investigations of the WEP
which are about as accurate as LLR
\cite{Baessler_etal_1999,Adelberger_2001}. LLR is the dominant
test of the SEP with the most accurate testing of the EP at the
level of $\Delta (M_G/M_I)_{EP} =(-1.0\pm1.4)\times10^{-13}$
\cite{Williams_Turyshev_Boggs_2004}. This result corresponds to a
test of the SEP of $\Delta (M_G/M_I)_{SEP}
=(-2.0\pm2.0)\times10^{-13}$ with the SEP violation parameter
$\eta=4\beta-\gamma-3$ found to be $\eta=(4.4\pm4.5)\times
10^{-4}$. Using the recent Cassini result for the PPN parameter
$\gamma$, PPN parameter $\beta$ is determined at the level of
$\beta-1=(1.2\pm1.1)\times 10^{-4}$ (see Figure~\ref{fig:ppn}).

%%%%%%%%%%%%%%%%%%%%%%%%%%%%%%%%%%%%%%%%%%%%%%%%%%%%%%%%%%%%%%
%%%%%%%%%%%%%%%%%%%%%%%%%%%%%%%%%%%%%%%%%%%%%%%%%%%%%%%%%%%%%%%

LLR data yielded the strongest limits to date on variability of
the gravitational constant (the way gravity is affected by the
expansion of the Universe), the best measurement of the de Sitter
precession rate, and is relied upon to generate accurate
astronomical ephemerides. The possibility of a time variation of
the gravitational constant, {\it G}, was first considered by Dirac
in 1938 on the basis of his large number hypothesis, and later
developed by Brans and Dicke in their theory of gravitation (for
more details consult \cite{Will1,Will_2001}). Variation might be
related to the expansion of the Universe, in which case $\dot
G/G=\sigma H_0$, where $H_0 $ is the Hubble constant, and $\sigma$
is a dimensionless parameter whose value depends on both the
gravitational constant and the cosmological model considered.
Revival of interest in Brans-Dicke-like theories, with a variable
{\it G}, was partially motivated by the appearance of string
theories where {\it G} is considered to be a dynamical quantity
\cite{Marciano1984}.

In Brans-Dicke theory, as well as in more general scalar-tensor
theories, the gravitational coupling depends on the cosmic time.
Observational bounds arising from the timing of the binary pulsar
PSR1913+16 yield quite restrictive bounds \cite{Damour1} of $
\dot{G} / G = (1.0 \pm 2.3) \times 10^{-11}~{\rm yr^{-1}}$, with a
magnitude similar to the cosmological bounds available
\cite{Bertolami8,Gillies,Chiba} (see Ref. \cite{Bertolami13} and
references therein for a discussion on a connection with the
accelerated expansion of the Universe). Varying-G solar models
\cite{Guenther} and measurements of masses and ages of neutron
stars yield even more stringent limits \cite{Thorsett} of $
\dot{G} / G = (- 0.6 \pm 2.0) \times 10^{-12}~{\rm yr^{-1}}$.

The most stringent limit on a change of $G$ comes from LLR, which
is one of the important gravitational physics result that LLR
provides. GR does not predict a changing $G$, but
some other theories do, thus testing for this effect is important.
As we have seen, the most accurate limit published is the current
LLR test, yielding $ \dot{G} / G = (4 \pm 9) \times 10^{-13}~{\rm
yr^{-1}}$ \cite{Williams_Turyshev_Boggs_2004}. The $\dot{G}/G$
uncertainty is 83 times smaller than the inverse age of the
Universe, $t_0=13.4$~Gyr with the value for Hubble constant
$H_0=72$~km/sec/Mpc from the WMAP data \cite{Spergel:2003cb}. The
uncertainty for $\dot G/G$ is improving rapidly because its
sensitivity depends on the square of the data span. This fact puts
LLR, with its more then 35 years of history, in a clear advantage
as opposed to other experiments.

LLR has also provided the only accurate determination of the
geodetic precession. Ref. \cite{Williams_Turyshev_Boggs_2004}
reports a test of geodetic precession, which expressed as a
relative deviation from GR, is $K_{gp}=-0.0019\pm0.0064$. The GP-B
satellite should provide improved accuracy over this value, if
that mission is successfully completed. LLR also has the
capability of determining PPN $\beta$ and $\gamma$ directly from
the point-mass orbit perturbations. A future possibility is
detection of the solar $J_2$ from LLR data combined with the
planetary ranging data. Also possible are dark matter tests,
looking for any departure from the inverse square law of gravity,
and checking for a variation of the speed of light. The accurate
LLR data has been able to quickly eliminate several suggested
alterations of physical laws. The precisely measured lunar motion
is a reality that any proposed laws of attraction and motion must
satisfy.

%%%%%%%%%%%%%%%%%%%%%%%%%%%% Section 3 %%%%%%%%%%%%%%%%%%%%%%%%%%%%%%%%%
\section{Search for New Physics Beyond General Relativity}
\label{sec:beyond}

The nature of gravity is fundamental to the understanding of the
solar system and the large scale structure of the Universe. This
importance motivates various precision tests of gravity both in
laboratories and in space.  To date, the experimental evidence for
gravitational physics is in agreement with GR; however, there are
a number of reasons to question the validity of this theory.
Despite the success of modern gauge field theories in describing
the electromagnetic, weak, and strong interactions, it is still
not understood how gravity should be described at the quantum
level. In theories that attempt to include gravity, new long-range
forces can arise in addition to the Newtonian inverse-square law.
Even at the purely classical level, and assuming the validity of
the Equivalence Principle, Einstein's theory does not provide the
most general way to describe the space-time dynamics, as there are
reasons to consider additional fields and, in particular, scalar
fields.

Although scalar fields naturally appear in the modern theories,
their inclusion predicts a non-Einsteinian behavior of gravitating
systems.  These deviations from GR lead to a
violation of the EP, modification of large-scale gravitational
phenomena, and imply that the constancy of the ``constants'' must
be reconsidered. These predictions motivate searches for small
deviations of relativistic gravity from GR and provide a
theoretical paradigm and constructive guidance for further gravity
experiments.  As a result, this theoretical progress has motivated
high precision tests of relativistic gravity and especially those
searching for a possible violation of the Equivalence Principle.
Moreover, because of the ever increasing practical significance of
the general theory of relativity (i.e. its use in spacecraft
navigation, time transfer, clock synchronization, standards of
time, weight and length, \textit{etc}.) this fundamental theory must be
tested to increasing accuracy.

An understanding of gravity at a quantum level will allow one to
ascertain whether the gravitational ``constant'' is a running
coupling constant like those of other fundamental interactions of
Nature. String/M-theory \cite{Green} hints a negative answer to
this question, given the non-renormalization theorems of
Supersymmetry, a symmetry at the core of the underlying principle
of string/M-theory and brane models, \cite{Polchinski,Randall}.
However, 1-loop higher--derivative quantum gravity models may
permit a running gravitational coupling, as these models are
asymptotically free, a striking property \cite{Julve}. In the
absence of a screening mechanism for gravity, asymptotic freedom
may imply that quantum gravitational corrections take effect on
macroscopic and even cosmological scales, which of course has some
bearing on the dark matter problem \cite{Goldman} and, in
particular, on the subject of the large scale structure of the
Universe \cite{Bertolami11,Bertolami12} (see, however,
\cite{Bertolami8}). Either way, it seems plausible to assume that
quantum gravity effects manifest themselves only on cosmological
scales.

In this Section we review the arguments for high-accuracy
experiments motivated by the ideas suggested by proposals of
quantization of gravity.

\subsection{String/M-Theory}
\label{sec:SMT}

String theory is currently referred to as string/M-theory, given
the unification of the existing string theories that is achieved 
in the context M-theory. Nowadays, it is widely viewed as the most
promising scheme to make GR compatible with
quantum mechanics (see \cite{Green} for an extensive
presentation). The closed string theory has a spectrum that contains as
zero mass eigenstates the graviton, $g_{MN}$, the dilaton, $\Phi$,
and an antisymmetric second-order tensor, $B_{MN}$. There are
various ways in which to extract the physics of our
four-dimensional world, and a major difficulty lies in finding a
natural mechanism that fixes the value of the dilaton field, since
it does not acquire a potential at any order in string
perturbation theory.

Damour and Polyakov \cite{Damour_Polyakov_1994a} have studied a
possible a mechanism to circumvent the above difficulty by
suggesting string loop-contributions, which are counted by dilaton
interactions, instead of a potential. After dropping the
antisymmetric second-order tensor and introducing fermions, $\hat
\psi$, Yang-Mills fields, $\hat A^{\mu}$, with field strength
$\hat F_{\mu \nu}$, in a spacetime described by the metric $\hat
g_{\mu \nu}$, the relevant effective low-energy four-dimensional
action is

\beq S = \int_M d^4 x \sqrt{-\hat g} B(\Phi)\left[{1 \over
\alpha^{\prime}} [\hat R + 4 \hat \nabla_{\mu} \hat \nabla^{\mu}
\Phi - 4 (\hat \nabla \Phi)^2] - {k \over 4} \hat F_{\mu \nu} \hat
F^{\mu \nu} - \overline{\hat \psi} \gamma^{\mu} \hat D_{\mu} \hat
\psi + ... \right], \label{eq:3.1} \eeq

\noindent where

\beq B(\Phi) = e^{-2 \Phi} + c_0 + c_1 e^{2 \Phi} + c_2 e^{4 \Phi}
+ ..., \label{eq:3.2} \eeq

\noindent $\alpha^{\prime}$ is the inverse of the string tension
and $k$ is a gauge group constant; the constants $c_0$, $c_1$,
..., \textit{etc}., can, in principle, be determined via computation.

In order to recover Einsteinian gravity, a conformal
transformation must be performed with $ g_{\mu \nu} = B(\Phi) \hat
g_{\mu \nu}$, leading to an action where the coupling constants
and masses are functions of the rescaled dilaton, $\phi$:

\beq S = \int_M d^4 x \sqrt{-g} \left[{1 \over 4q} R - {1 \over
2q} (\nabla \phi)^2 - {k \over 4} B(\phi) F_{\mu \nu} F^{\mu \nu}
- \overline{\psi} \gamma^{\mu} D_{\mu} \psi + ... \right],
\label{eq:3.4} \eeq

\noindent from which follows that $4q = 16 \pi G = {1 \over 4}
\alpha^{\prime}$ and the coupling constants and masses are now
dilaton-dependent, through $ g^{-2} = k B(\phi)$ and $ m_A =
m_A(B(\phi))$. Damour and Polyakov proposed the minimal coupling
principle (MCP), stating that the dilaton is dynamically driven
towards a local minimum of all masses (corresponding to a local
maximum of {$B(\phi)$). Due to the MCP, the dependence of the
masses on the dilaton implies that particles fall differently in a
gravitational field, and hence are in violation of the WEP.
Although, in the solar system conditions, the effect is rather
small  being of the order of $\Delta a / a \simeq 10^{-18}$,
application of already available technology can potentially test
prediction. Verifying this prediction is an interesting prospect,
as it would present a distinct experimental signature of
string/M-theory. We have no doubts that the experimental search
for violations of the WEP, as well as of the fundamental Lorentz
and CPT symmetries, present important windows of opportunity to
string physics and should be vigorously pursued.

Recent analysis of a potential scalar field's evolution scenario
based on action (\ref{eq:3.4}) discovered that the present
agreement between GR and experiment might be naturally compatible
with the existence of a scalar contribution to gravity. In
particular, Damour and Nordtvedt
\cite{Damour_Nordtvedt_1993a} (see also
\cite{Damour_Polyakov_1994a} for non-metric versions
of this mechanism together with \cite{DPV02a} for the
recent summary of a dilaton-runaway scenario) have found that a
scalar-tensor theory of gravity may contain a ``built-in''
cosmological attractor mechanism toward GR. These scenarios assume
that the scalar coupling parameter $\frac{1}{2}(1-\gamma)$ was of
order one in the early Universe (say, before inflation), and show
that it then evolves to be close to, but not exactly equal to,
zero at the present time.

The Eddington parameter $\gamma$, whose value in general
relativity is unity, is perhaps the most fundamental PPN
parameter, in that $\frac{1}{2}(1-\gamma)$ is a measure, for
example, of the fractional strength of the scalar gravity
interaction in scalar-tensor theories of gravity
\cite{Damour_EFarese96a,Damour_EFarese96b}.  Within perturbation
theory for such theories, all other PPN parameters to all
relativistic orders collapse to their general relativistic values
in proportion to $\frac{1}{2}(1-\gamma)$. Under some assumptions
(see \textit{e.g.} \cite{Damour_Nordtvedt_1993a}) one
can even estimate what is the likely order of magnitude of the
left-over coupling strength at present time which, depending on
the total mass density of the Universe, can be given as $1-\gamma
\sim 7.3 \times 10^{-7}(H_0/\Omega_0^3)^{1/2}$, where $\Omega_0$
is the ratio of the current density to the closure density and
$H_0$ is the Hubble constant in units of $100~ km/sec/Mpc$. Compared
to the cosmological constant, these scalar field models are
consistent with the supernovae observations for a lower matter
density, $\Omega_0\sim 0.2$, and a higher age, $(H_0 t_0) \approx
1$. If this is indeed the case, the level $(1-\gamma) \sim
10^{-6}-10^{-7}$ would be the lower bound for the present value of
the PPN parameter $\gamma$
\cite{Damour_Nordtvedt_1993a}. This is why
measuring the parameter $\gamma$ to accuracy of one part in a
billion, as suggested for the LATOR mission
\cite{Laser_Clocks_LATOR}, is important.

\subsection{Scalar-Tensor Theories of Gravity} \label{sec:vacuum}

In many alternative theories of gravity, the gravitational
coupling strength exhibits a dependence on a field of some sort;
in scalar-tensor theories, this is a scalar field $\varphi$. A
general action for these theories can be written as
{}

\beq  S= {c^3\over 4\pi G}\int d^4x \sqrt{-g}
\left[\frac{1}{4}f(\varphi) R - \frac{1}{2}g(\varphi) \partial_{\mu} \varphi
\partial^{\mu} \varphi + V(\varphi) \right] + \sum_{i}
q_{i}(\varphi)\mathcal{L}_{i}, \label{eq:sc-tensor} \eeq
\noindent where $f(\varphi)$, $g(\varphi)$, $V(\varphi)$ are
generic functions, $q_i(\varphi)$ are coupling functions and
$\mathcal{L}_{i}$ is the Lagrangian density of the matter fields;
it is worth mentioning that the graviton-dilaton system in
string/M-theory can be viewed as one of such scalar-tensor
theories of gravity. An emblematic proposal is the well-known
Brans-Dicke theory \cite{Brans} corresponds to the specific choice

\beq \label{eq6:2} f(\varphi) = \varphi, \qquad g(\varphi) =
{\omega \over \varphi}, 
\eeq
and a vanishing potential $V(\varphi)$. Notice that in
the Brans-Dicke theory the kinetic energy term of the field
$\varphi$ is non-canonical, and the latter has a dimension of
energy squared. In this theory, the constant $\omega$ marks
observational deviations from GR, which is recovered in the limit
$\omega \to \infty$. We point out that, in the context of the
Brans-Dicke theory, one can operationally introduce the Mach's
Principle which, we recall, states that the inertia of bodies is
due to their interaction with the matter distribution in the
Universe. Indeed, in this theory the gravitational coupling is
proportional to $\varphi^{-1}$, which depends on the
energy-momentum tensor of matter through the field equations.
Observational bounds require that $|\omega| \gsim 500$
\cite{RoberstonCarter91,viking_reasen}, and even higher values $|\omega| \gsim 40000$
are reported in \cite{Will2005}. In the so-called {\it induced gravity models} \cite{Fujii}, the
functions of the fields are initially given by $ f(\varphi) =
\varphi^2$ and $ g(\varphi) = 1 / 2$, and the potential
$V(\varphi)$ allows for a spontaneous symmetry breaking, so that
the field $\varphi$ acquires a non-vanishing vacuum expectation
value, $f(\vev{0|\varphi|0}) = \vev{0|\varphi^2|0} = M_P^2 =
G^{-1}$. Naturally the cosmological constant is given by the
interplay of the value $V(\vev{0|\varphi|0})$ and all other
contributions to the vacuum energy.

Therefore, it is clear that in this setup Newton's constant arises
from dynamical or symmetry-breaking considerations. It is
mesmerizing to conjecture that the $\varphi$ field could be
locally altered: this would require the coupling of this field
with other fields, in order to locally modify its value. This
feature can be found in some adjusting mechanisms devised as a
solution of the cosmological constant problem (see \textit{e.g.}
\cite{Weinberg1} for a list of references). However, Weinberg
\cite{Weinberg1} has shown that these mechanisms are actually
unsuitable for this purpose, although they nevertheless contain
interesting multi-field dynamics. Recent speculations suggesting
that extra dimensions in braneworld scenarios may be rather large
\cite{Antoniadis,Arkani} bring forth gravitational effects at the
much lower scale set by $M_5$, the 5-dimensional Planck mass.
Phenomenologically, the existence of extra dimensions should
manifest itself through a contribution to Newton's law on small
scales, $r \lsim 10^{-4}~m$, as discussed next in Section
\ref{sec:new-inter}.

\subsection{Search for New Interactions of Nature}
\label{sec:new-inter}

The existence of new fundamental forces beyond the already known
four fundamental interactions, if confirmed, will have several
implications and bring important insights into the physics beyond
the Standard Model. A great interest on the subject was sparked
after the 1986 claim of evidence for an intermediate range
interaction with sub-gravitational strength \cite{Fishbach1}, both
theoretical (see \cite{Nieto} for a review) as well as
experimental, giving rise to a wave of new setups, as well as
repetitions of ``classical'' ones using state of the art
technology.

In its simplest versions, a putative new interaction or a fifth
force would arise from the exchange of a light boson coupled to
matter with a strength comparable to gravity. Planck-scale physics
could give origin to such an interaction in a variety of ways,
thus yielding a Yukawa-type modification in the interaction energy
between point-like masses. This new interaction can be derived,
for instance, from extended supergravity theories after
dimensional reduction \cite{Nieto,Scherk}, compactification of
$5$-dimensional generalized Kaluza-Klein theories including gauge
interactions at higher dimensions \cite{Bars}, and also from
string/M-theory. In general, the interaction energy, $V(r)$,
between two point masses $m_1$ and $m_2$ can be expressed in terms
of the gravitational interaction as\footnote{We use here the units $c =
\hbar = 1$.}
\beq V(r) = -  {G_{\infty}m_{1}m_{2} \over r}\big(1 +
\alpha\,e^{-r/\lambda}\big), \label{eq:2.1} \eeq

\noindent where $r = \vert {\bf r}_2 - {\bf r}_1 \vert$ is the
distance between the masses, $G_{\infty}$ is the gravitational
coupling for $r \rightarrow \infty$ and $\alpha$ and $\lambda$ are
respectively the strength and range of the new interaction.
Naturally, $G_{\infty}$ has to be identified with Newton's
gravitational constant and the gravitational coupling becomes
dependent on $r$. Indeed, the force associated with Eq.
(\ref{eq:2.1}) is given by:
\beq {\bf F}({\bf r}) = - \nabla V({\bf r}) = - {G(r)m_{1}m_{2}
\over r^2}\,\hat {\bf r}, \label{eq:2.2} \eeq
\noindent where
\beq G(r) = G_{\infty}\big[1 + \alpha\,(1 +
r/\lambda)e^{-r/\lambda}\big]. \label{eq:2.3} \eeq
\noindent The suggestion of existence of a new interaction arose
from assuming that the coupling $\alpha$ is not an universal
constant, but instead a parameter depending on the chemical
composition of the test masses \cite{Lee}. This comes about if one
considers that the new bosonic field couples to the baryon number
$B = Z + N$, which is the sum of protons and neutrons. Hence the
new interaction between masses with baryon numbers $B_1$ and $B_2$
can be expressed through a new fundamental constant, $f$, as:
\beq V(r) = - f^{2}{B_{1}B_{2} \over r}e^{-r/\lambda},
\label{eq:2.4} \eeq
\noindent such that the constant $\alpha$ can be written as
\beq \alpha= - \sigma \left({B_{1} \over \mu_{1}}\right)
\left({B_{2} \over \mu_{2}}\right), \label{eq:2.5} \eeq
\noindent with $\sigma = f^{2}/G_{\infty}m_{H}^{2}$ and $\mu_{1,2}
= m_{1,2}/m_{H}$, $m_H$ being the hydrogen mass.

The above equations imply that in a Galileo-type experiment a
difference in acceleration exists between the masses $m_1$ and
$m_2$, given by
\beq {\bf a}_{12}= \sigma\left({B \over \mu}\right)_\oplus
\left[\left({B_{1} \over \mu_{1}}\right)- \left({B_{2} \over
\mu_{2}}\right)\right]{\bf g}, \label{eq:2.6} \eeq
\noindent where {$\bf g$} is the field strength of the Earth's
gravitational field.

Several experiments (see, for instance, Refs.
\cite{Fishbach1,Nieto} for a list of the most relevant) studied
the parameters of a new interaction based on the idea of a
composition-dependence differential acceleration, as described in
Eq.~(\ref{eq:2.6}), and other composition-independent
effect\footnote{For instance, neutron interferometry has been
suggested to investigate a possible new force that couples to
neutron number \cite{Bertolami10}.}. The current experimental
status is essentially compatible with the predictions of Newtonian
gravity, in both composition-independent or -dependent setups. The
bounds on parameters $\alpha$ and $\lambda$ are summarized below
(Figure~\ref{fig:Figure 1}):
\begin{itemize}
\item[--] Laboratory experiments devised to measure deviations
from the inverse-square law are most sensitive in the range
$10^{-2}~{\rm m} \lsim \lambda \lsim 1~{\rm m}$, constraining
$\alpha$ to be smaller than about $10^{-4}$;

\item[--] Gravimetric experiments sensitive in the range of
$10~{\rm m} \lsim \lambda \lsim 10^{3}~{\rm m}$ indicate that
$\alpha \lsim 10^{-3}$;

\item[--] Satellite tests probe the ranges of about 
$10^{5}~{\rm m} \lsim \lambda \lsim 10^{7}~{\rm m}$ suggest 
that $\alpha \lsim 10^{-7}$;

\item[--] Analysis of the effects of the inclusion of scalar
fields into the stellar structure yields a bound in the range
$10^8~{\rm m} \lsim \lambda \lsim 10^{10}~{\rm m}$, limiting
$\alpha$ to be smaller than approximately $10^{-2}$
\cite{Bertolami-Paramos2005b}.
\end{itemize}
\noindent The latter bound, although modest, is derived from a
simple computation of the stellar equilibrium configuration in the
polytropic gas approximation, when an extra force due to a Yukawa
potential is taken into account on the hydrostatic equilibrium
equation.

%%%%%%%%%%%%%%%%%%%%%%%%%%%%%%%%%%%%%%%%%%%%%%%%%%%%%%%%%%%%%%
\begin{figure}[t]
\centering
\leavevmode\epsfysize=9.5cm \epsfbox{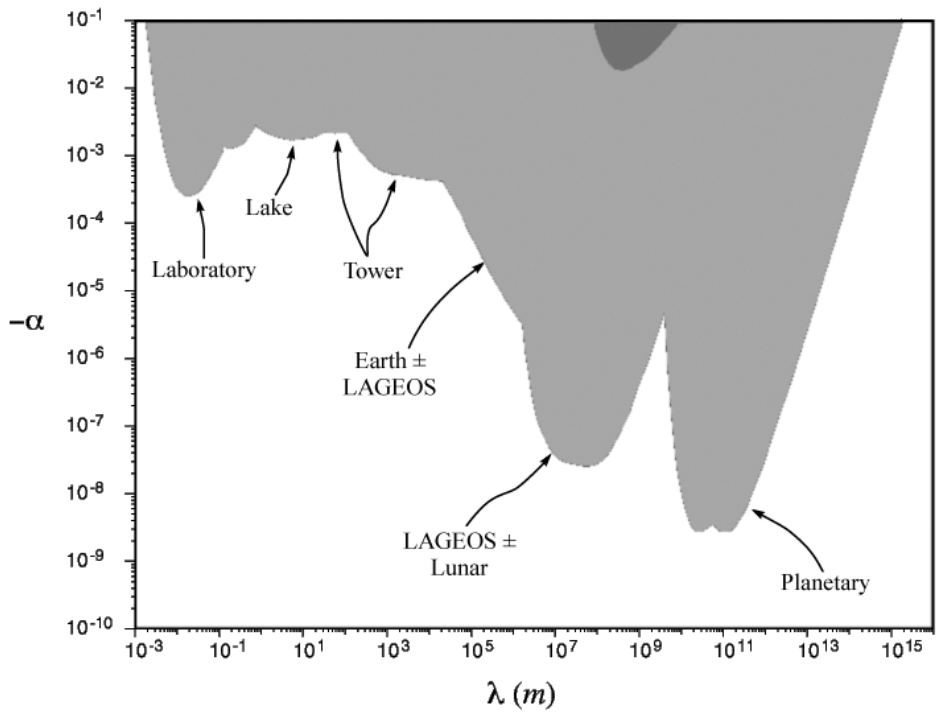}\\
\caption{Experimentally excluded regions for the range and
strength of possible new forces, as shown in Ref.
\cite{Bertolami-Paramos2005b}.} \label{fig:Figure 1}
\end{figure}
%%%%%%%%%%%%%%%%%%%%%%%%%%%%%%%%%%%%%%%%%%%%%%%%%%%%%%%%%%%%%%%

Remarkably, $\alpha$ is so far essentially unconstrained for
$\lambda < 10^{-3}~{\rm m}$ and $\lambda > 10^{13}~{\rm m}$. The
former range is particularly attractive as a testing ground for
new interactions, since forces with sub-millimetric range seems to
be favored from scalar interactions in supersymmetric theories
\cite{Dimopoulos}; this is also the case in the recently proposed
theories of $\tev$ scale quantum gravity, which stem from the
hypothesis that extra dimensions are not necessarily of Planck
size \cite{Antoniadis,Arkani}. The range $\lambda < 10^{-3}~{\rm
m}$ also arises if one assumes that scalar \cite{Beane} or tensor
interactions associated with Lorentz symmetry breaking in string
theories \cite{Bertolami3} account for the vacuum energy up to
half of the critical density. Putative
corrections to Newton's law at millimeter range could have
relevant implications, especially taking into account that, in
certain models of extra dimensions, these corrections can be as
important as the usual Newtonian gravity \cite{Arkani,Floratos}. 
From the experimental side, this range has recently been available
to experimental verification; state of the art experiments rule
out extra dimensions over length scales down to $0.2$~mm
\cite{Hoyle}.

\subsection{Gravity Shielding - the Majorana Effect}
\label{majorana}

The possibility that matter can shield gravity is not predicted by
modern theories of gravity, but it is a recurrent idea and it
would cause a violation of the equivalence principle test. In
fact, the topic has been more recently reviewed in \cite{Gillies}
renewing the legitimacy of this controversial proposal;
consequently, a brief discussion is given in this subsection.

The idea of gravity shielding goes back at least as far as to
Majorana's 1920 paper \cite{Majorana_1920}. Since then a number of
proposals and studies has been put forward and performed to test
the possible absorption of the gravitational force between two
bodies when screened from each other by a medium other than
vacuum.  This effect is a clear gravitational analog of the
magnetic permeability of materials, and Majorana
\cite{Majorana_1920} suggested the introduction of a screening or
extinction coefficient, $h$, in order to measure the shielding of
the gravitational force between masses $m_1$ and $m_2$ induced by
a material with density $\rho(r)$; such an effect can be modeled
as

\beq F^{\prime} = {G \,m_1 m_2 \over r^2} \exp\left[- h \int
\rho(r)~dr \right]. \label{eq:2.10} 
\eeq
which clearly depends on the amount of mass between
attracting mass elements and a universal constant $h$. Naturally,
one expects $h$ to be quite small.

Several attempts to derive this parameter from general principles
have been made. Majorana gave a closed form expression for a
sphere's gravitational to inertial mass ratio. For weak shielding
a simpler expression is given by the linear expansion of the
exponential term, $M_G / M_I \approx 1 - h f R \bar{\rho}$, where
$f$ is a numerical factor, $\bar{\rho}$ is the mean density, and
$R$ is the sphere's radius. For a homogeneous sphere Majorana and
Russell give $f=3/4$. For a radial density distribution of the
form $\rho(r)=\rho(0) (1-r^2/R^2)^n$ Russell derives
$f=(2n+3)^2/(12n+12)$.  Russell \cite{Russell_1921} realized that
the large masses of the Earth, Moon and planets made the
observations of the orbits of these bodies and the asteroid Eros a
good test of such a possibility. He made a rough estimate that the
equivalence principle was satisfied to a few parts per million,
which was much smaller than a numerical prediction based on
Majorana's estimate for $h$. If mass shields gravity, then large
bodies such as, for instance,  the Moon and Earth will partly
shield their own gravitational attraction. The observable ratio of
gravitational mass to inertial mass would not be independent of
mass, which would violate the equivalence principle.

Eckhardt \cite{Eckhardt}  showed that LLR can be used to set the
limit $h \leq 1.0 \times 10^{-22}~{\rm m^2~kg^{-1}}$, six orders
of magnitude smaller than the geophysical constraint. In
\cite{Eckhardt}, an LLR test of the equivalence principle was used
to set a modern limit on gravity shielding. That result is updated
as follows: the uniform density approximation is sufficient for
the Moon and $f\, R \,\bar{\rho} = 4.4 \times 10^9$ kg~m$^{-2}$. 
For the Earth we use $n\approx 0.8$ with Russell's expression to get
$f\, R\,\bar{\rho} = 3.4 \times10^{10}$ kg~m$^{-2}$. Using the
difference $-3.0 \times 10^9~ g/cm^2$ $h$ along with the LLR EP
solution for the difference in gravitational to inertial mass
ratios gives $h = (3 \pm 5) \times 10^{-24}$ m$^2$~kg$^{-1}$
\cite{pescara05}. The value is not significant compared to the
uncertainty. To give a sense of scale to the uncertainty, for the
gravitational attraction to be diminished by $1/2$ would require a
column of matter with the density of water stretching at least
half way from the solar system to the center of the galaxy. The
LLR equivalence principle tests give no evidence that mass shields
gravity and the limits are very strong.

For completeness, let us mention that Weber \cite{Weber} argued
that a quasi-static shielding could be predicted from a analysis
of relativistic tidal phenomena, concluding that such effect
should be extremely small. Finally, the most stringent laboratory
limit on the gravitational shielding constant had been obtained
during the recent measurement of Newton's constant, resulting in
$h \leq 4.3 \times 10^{-15}~{\rm m^2~kg^{-1}}$
\cite{Unnikrishnan}.

%%%%%%%%%%%%%%%%%%%%%%%%%% Section 4 %%%%%%%%%%%%%%%%%%%%%%%%%%%%%
\section{The ``Dark Side'' of Modern Physics}
\label{cosmological}

To a worldwide notice, recent cosmological observations dealt us a
challenging puzzle forcing us to accept the fact that our current
understanding of the origin and evolution of the Universe is incomplete. 
Surprisingly, it turns out that most of the energy content of the Universe 
is in the form of the presently unknown dark matter and dark energy that may 
likely permeate all of spacetime. It is possible that the underlying physics 
that resolve the discord between quantum mechanics and GR will 
also shed light on cosmological questions addressing the origin and ultimate 
destiny of the Universe.

In this Section we shall consider mechanisms that involve new physics
beyond GR to explain the puzzling behavior observed  at
galactic and cosmological scales.

\subsection{Cold Dark Matter}

The relative importance of the gravitational interaction increases
as one considers large scales, and it is at the largest scales
where the observed gravitational phenomena do not agree with our
expectations. Thus, based on the motion of the peripheral galaxies
of the Coma cluster of galaxies, in 1933 Fritz Zwicky found a
discrepancy between the value inferred from the total number of
galaxies and brightness of the cluster. Specifically, this
estimates of the total amount of mass in the cluster revealed the
need for about 400 times more mass than expected. This led Zwicky
conclude that there is another form of matter in the cluster
which, although unaccounted, contains most of the mass responsible
for the gravitational stability of the cluster. This non-luminous
matter became known as the ``dark matter''. The dark matter
hypothesis was further supported by related problems, namely the
differential rotation of our galaxy, as first discussed by Oort in
1927, and the flatness of galactic rotation curves \cite{Trimble}.

The most common approach to these problems is to assume the
presence of unseen forms of energy that bring into agreement the
observed phenomena with GR. The standard scenario to explain the
dynamics of galaxies consists in the introduction of an extra
weakly interacting massive particle, the so-called Cold Dark
Matter (CDM), that clusters at the scales of galaxies and provides
the required gravitational pull to hold them together. The
explanation of the observed acceleration of the expansion of the
Universe requires however the introduction of a more exotic form
of energy, not necessarily associated with any form of matter but
associated with the existence of space-time itself -- vacuum
energy.

Although CDM can be regarded as a natural possibility given our
knowledge of elementary particle theory, the existence of a
non-vanishing but very small vacuum energy remains an unsolved puzzle
for our high-energy understanding of physics. However, the CDM
hypothesis finds problems when one begin to look at the details
of the observations. Increasingly precise simulations of galaxy
formation and evolution, although relatively successful in broad
terms, show well-known features that seem at odds with their real
counterparts, the most prominent of which might be the ``cuspy
core'' problem and the over-abundance of substructure seen in the
simulations (see, for instance, \cite{Ostriker:2003qj}).

At the same time the CDM hypothesis is required to explain the
correlations of the relative abundances of dark and luminous
matter that seem to hold in a very diverse set of astrophysical
objects \cite{McGaugh:2005er}. These correlations are exemplified
in the Tully-Fisher law \cite{Tully:1977fu} and can be interpreted
as pointing to an underlying acceleration scale,
$a_0 \simeq 10^{-10}~m~s^{-2}$, below which the
Newtonian potential changes and gravity becomes stronger. This is
the basic idea of MOND (MOdified Newtonian Dynamics), a successful
phenomenological modification of Newton's potential proposed in
1983 \cite{Milgrom:1983ca} whose predictions for the rotation
curves of spiral galaxies have been realized with increasing
accuracy as the quality of the data has improved
\cite{Sanders:2002pf}. Interestingly, the critical acceleration 
required by the data is
of order $a_0 \sim c H_0$ where $H_0$ is today's Hubble constant
and $c$ the speed of light (that we will set to 1 from now on).
The problem with this idea is that MOND is just a modification of
Newton's potential so it is inadequate in any situation in which
relativistic effects are important. Efforts have been made to
obtain MONDian phenomenology in a relativistic generally covariant
theory by including other fields in the action with suitable
couplings to the spacetime metric \cite{Bekenstein:2004ne}.

On the other hand, in what concerns the CDM model one can state that 
if all matter is purely baryonic, early structure formation
does not occur, as its temperature and pressure could not account
for the latter. The presence of cold (i.e., non-relativistic) dark
matter allows for gravitational collapse and thus solves this
issue. Another hint of the existence of exotic dark matter lies in
the observation of gravitational lensing, which may be interpreted
as due to the presence of undetected clouds of non-luminous matter
between the emitting light source and us, which bends the light
path due to its mass. This could also be the cause for the
discrepancies in the measured Lyman-alpha forest, the spectra of
absorption lines of distant galaxies and quasars. The most likely
candidates to account for the dark matter include a linear
combination of neutral supersymmetrical particles, the neutralinos
(see \textit{e.g.} \cite{Munoz}), axions \cite{Asztalos},
self-interacting scalar particles \cite{Bento1}, \textit{etc}..

On a broader sense, one can say that
these models do not address in a unified way the Dark Energy
(discussed in Section~\ref{sec:de}) and Dark Matter problems,
while a common origin is suggested by the observed coincidence
between the critical acceleration scale and the Dark Energy
density. This unification feature is found in the so-called 
generalized Chaplygin gas {\cite{Bento2002} (see Section~\ref{sec:de} 
below)

\subsubsection{Modified Gravity as an Alternative to Dark Matter}

\label{sec:dm-grav-mod}

There are two types of effects in the dark-matter-inspired gravity
theories that are responsible for the infrared modification.
First, there is an extra scalar excitation of the spacetime metric
besides the massless graviton. The mass of this scalar field is of
the order of the Hubble scale in vacuum, but its mass depends
crucially on the background over which it propagates. This
dependence is such that this excitation becomes more massive near
the source, and the extra degree of freedom decouples at short
distances in the spacetime of a spherically symmetric mass. This
feature makes this excitation to behave in a way that remind of
the chameleon field of \cite{Khoury_Weltman_04,Khoury:2003aq,Brax_etal_04}, however, quite often
this ``chameleon'' field is just a component of the spacetime
metric coupled to the curvature.

There is also another effect in these theories -- the Planck mass
that controls the coupling strength of the massless graviton also
undergoes a rescaling or ``running'' with the distance to the
sources (or the background curvature). This phenomenon, although a
purely classical one in our theory, is reminiscent of the quantum
renormalization group running of couplings. So one might wonder if
MONDian type actions could be an effective classical description
of strong renormalization effects in the infrared that might
appear in GR \cite{Julve,Reuter:1996cp}, as happens in QCD. A
phenomenological approach to structure formation bored on these
effects has been attempted in Ref. \cite{Bertolami11}. Other
implications, such as lensing, cosmic virial theorem and
nucleosynthesis, were analyzed in Refs.
\cite{Bertolami8,Bertolami21, Bertolami22}. Additionally, these
models offer a phenomenology that seems well suited to describe an
infrared strongly coupled phase of gravity and especially at high
energies/curvatures when one may use the GR action or its
linearization being a good approximation; however, when one
approaches low energies/curvatures one finds a non-perturbative
regime. At even lower energies/curvatures perturbation theory is
again applicable, but the relevant theory is of scalar-tensor type
in a de Sitter space.

Clearly there are many modifications of the proposed class of
actions that would offer a similar phenomenology, such that
gravity would be modified below a characteristic acceleration
scale of the order of the one required in MOND. Many of these
theories also offer the unique possibility of being tested not
only through astrophysical observations, but also through
well-controlled laboratory experiments where the outcome of an
experiment is correlated with parameters that can be determined by
means of cosmological and astrophysical measurements.

\subsection{Dark Energy as Modern Cosmological Puzzle}
\label{sec:de}

In 1998 Perlmutter and collaborators \cite{Perlmutter} and Riess
and collaborators \cite{Riess} have gathered data of the
magnitude-redshift relation of Type Ia supernovae with redshifts
$z\ge 0.35$ and concluded that it strongly suggest that we live in
an accelerating Universe, with a low matter density with about one
third of the of the energy content of the Universe. Currently
there are about 250 supernovae data points which confirm this
interpretation. Dark energy is assumed to be a smooth distribution
of non-luminous energy uniformly distributed over the Universe so
to account for the extra dimming of the light of far away Type Ia
supernovae, standard candles for cosmological purposes.  If there
is a real physical field responsible for Dark Energy, it may be
phenomenologically described in terms of an energy density $\rho$
and pressure $p$, related instantaneously by the equation-of-state
parameter $w=p/\rho$. Furthermore, covariant energy conservation
would then imply that $\rho$ dilutes as $a^{-3(1+w)}$, with $a$
being the scale factor.  Note that $p=w\rho$ is not necessarily
the actual equation of state of the Dark Energy fluid, meaning
that perturbations may not generally obey $\delta p = w \delta
\rho$; however, if one were to have such an equation of state, one
can define the speed of sound by $c_{s}^{2} = \partial p/ \partial
\rho$.  The implications of this phenomenology would make much
more sense in the context of theories proposed to provide the
required microscopic description.

\subsubsection{Cosmological Constant and Dark Energy}

One of the leading explanations for the accelerated expansion of
the Universe is the presence of a non-zero cosmological constant.
As can be seen from Einstein's equation, the cosmological term can
be viewed not as a geometric prior to the spacetime continuum, but
instead interpreted as a energy-momentum tensor proportional to
the metric, thus enabling the search for the fundamental physics
mechanism behind its value and, possibly, its evolution with
cosmic time. An outstanding question in today's physics lies in
the discrepancy between the observed value for $\La$ and the
prediction arising from quantum field theory, which yields a
vacuum energy density about $120$ orders of magnitude larger than
the former. To match the observed value, requires a yet unknown
cancelation mechanism to circumvent the fine tuning of $120$
decimal places necessary to account for the observations. This is
so as observations require the cosmological constant to be of
order of the  critical density $\rho_c = 3H_0^2/8\pi G \simeq 
10^{-29}$~g~cm$^{-3}$:

\beq \rho_V \equiv {\Lambda \over 8 \pi G} \simeq 10^{-29}
{\rm g~cm^{-3}} \simeq 10^{-12} {\rm eV^4}, \label{eq:1.4} \eeq

\noindent while the natural number to expect from a quantum gravity theory 
is $M_P^4 \simeq 10^{76}$~GeV$^4$.

Besides the cosmological constant, a slow-varying vacuum 
energy\footnote{For earlier suggestions see Refs. \cite{Bronstein38}.} of
some scalar field, usually referred to as ``quintessence''
\cite{Caldwell}, or an exotic fluid like the generalized Chaplygin
gas {\cite{Bento2002} are among other the most discussed
candidates to account for this dominating contribution for the
energy density. It is worth mentioning that the latter possibility
allows for a scenario where dark energy and dark matter are
unified.

We mention that the presented bounds result from a variety of
sources, of which the most significant are the CMBR, 
high-$z$ supernovae redshifts and
galaxy cluster abundances. These joint constraints establish that
the amount of dark energy, dark matter and baryons are, in terms
of the critical density, $\Om_\La \simeq 0.73$, $\Om_{\tt DM} =
0.23$ and $\Om_{\tt Baryons} = 0.04$, respectively \cite{LahavLiddle05}.

Current observational constraints imply that the evolution of Dark
Energy is entirely consistent with $w=-1$, characteristic of a
cosmological constant ($\Lambda$). The cosmological constant was
the first, and remains the simplest, theoretical solution to the
Dark Energy observations.  The well-known ``cosmological constant
problem'' -- why is the vacuum energy so much smaller than we
expect from effective-field-theory considerations? -- remains, of
course, unsolved.

Recently an alternative mechanism to explain $\Lambda$ has arisen
out of string theory. It was previously widely perceived that
string theory would continue in the path of QED and QCD wherein
the theoretical picture contained few parameters and a uniquely
defined ground state. However recent developments have yielded a
theoretical horizon in distinct opposition to this, with a
``landscape'' of possible vacua generated during the
compactification of 11 dimensions down to 3 \cite{Kachru_KL_03}.
Given the complexity of the landscape, anthropic arguments have
been put forward to determine whether one vacuum is preferred over
another. It is possible that further development of the statistics
of the vacua distribution, and characterization of any distinctive
observational signatures, such as predictions for the other
fundamental coupling constants, might help to distinguish
preferred vacua and extend beyond the current vacua counting
approach.

Although Dark Energy is the most obvious and popular possibility
to the recently observed acceleration of the Universe, other
competing ideas have been investigated, and among them is
modifications of gravity on cosmological scales.  Indeed, as we
discussed earlier, GR is well tested in the solar system, in
measurements of the period of the binary pulsar, and in the early
Universe, via primordial nucleosynthesis. None of these tests,
however, probes the ultra-large length scales and low curvatures
characteristic of the Hubble radius today.  Therefore, one can
potentially think that gravity is modified in the very far
infrared allowing the Universe to accelerate at late times.

In this section we will discuses some of the gravity modification
proposals suggested to provide a description of the observed
acceleration of the Universe.

\subsubsection{Modified Gravity as an Alternative to Dark Energy}

A straightforward possibility is to modify the usual
Einstein-Hilbert action by adding terms that are blow up as the
scalar curvature goes to zero
\cite{Carroll_DTT_2003,Carroll_FDETT_2004}.  Recently, models
involving inverse powers of the curvature have been proposed as an
alternative to Dark Energy
\cite{Carroll_FDETT_2004,Capozziello:2003tk}. In these models one
generically has more propagating degrees of freedom in the
gravitational sector than the two contained in the massless
graviton in GR. The simplest models of this kind add inverse
powers of the scalar curvature to the action ($\Delta {\cal
L}\propto 1/R^n$), thereby introducing a new scalar excitation in
the spectrum. For the values of the parameters required to explain
the acceleration of the Universe this scalar field is almost
massless in vacuum and one might worry about the presence of a new
force contradicting solar system experiments. However, it can be
shown that models that involve inverse powers of other invariant,
in particular those that diverge for $r\rightarrow 0$ in the
Schwarzschild solution, generically recover an acceptable weak
field limit at short distances from sources by means of a
screening or shielding of the extra degrees of freedom at short
distances \cite{Navarro:2005da}. Such theories can lead to
late-time acceleration, but unfortunately typically lead to one of
two problems. Either they are in conflict with tests of GR in the
solar system, due to the existence of additional dynamical degrees
of freedom \cite{Chiba_2003}, or they contain ghost-like degrees
of freedom that seem difficult to reconcile with fundamental
theories.  The search is ongoing for versions of this idea that
are consistent with experiment.

A more dramatic approach would be to imagine that we live on a
brane embedded in a large extra dimension.  Although such theories
can lead to perfectly conventional gravity on large scales, it is
also possible to choose the dynamics in such a way that new
effects show up exclusively in the far infrared.  An example is
the Dvali-Gabadadze-Porrati (DGP) braneworld model, in which the
strength of gravity in the bulk is substantially less than that on
the brane \cite{Dvali_GP_2000}.  Such theories can naturally lead
to late-time acceleration \cite{Deffayet_2000,Deffayet_etal_2001},
but may have difficulties with strong-coupling issues
\cite{Luty_PR_2003}. Furthermore, the DGP model does not properly
account for the supernova data, as does its generalization, the
Dvali-Turner model, and also other \textit{ad hoc} modifications
of the Friedmann equation, the so-called Cardassian model
\cite{Bento2005}. Most interestingly, however, DGP gravity and
other modifications of GR hold out the possibility of having
interesting and testable predictions that distinguish them from
models of dynamical Dark Energy. One outcome of this work is that
the physics of the accelerating Universe may be deeply tied to the
properties of gravity on relatively short scales, from millimeters
to astronomical units.

\subsubsection{Scalar field Models as Candidate for Dark Energy}
\label{sec:sc-models-de}

One of the simplest candidates for dynamical Dark Energy is a
scalar field , $\varphi$, with an extremely low-mass and an
effective potential, $V(\varphi)$, as shown by
Eq.~(\ref{eq:sc-tensor}) \cite{Bertolami13}.  If the field is
rolling slowly, its persistent potential energy is responsible for
creating the late epoch of inflation we observe today. For the
models that include only inverse powers of the curvature, besides
the Einstein-Hilbert term, it is however possible that in regions
where the curvature is large the scalar has naturally a large mass
and this could make the dynamics to be similar to those of GR
\cite{Cembranos:2005fi}. At the same time, the scalar curvature,
while being larger than its mean cosmological value, is still very
small in the solar system (to satisfy the available results of
gravitational tests). Although a rigorous quantitative analysis of
the predictions of these models for the tests in the solar system
is still noticeably missing in the literature, it is not clear
whether these models may be regarded as a viable alternative to
Dark Energy.

Effective scalar fields are prevalent in supersymmetric field
theories and string/M-theory. For example, string theory predicts
that the vacuum expectation value of a scalar field, the dilaton,
determines the relationship between the gauge and gravitational
couplings. A general, low energy effective action for the massless
modes of the dilaton can be cast as a scalar-tensor theory as
Eq.~(\ref{eq:sc-tensor}) with a vanishing potential, where
$f(\varphi)$, $g(\varphi)$ and $q_{i}(\varphi)$ are the dilatonic
couplings to gravity, the scalar kinetic term and gauge and matter
fields respectively, encoding the effects of loop effects and
potentially non-perturbative corrections.

A string-scale cosmological constant or exponential dilaton
potential in the string frame translates into an exponential
potential in the Einstein frame. Such quintessence potentials
\cite{Wetterich_88,Ratra_Peebles_88} can have scaling
\cite{Ferreira_Joyce_97}, and tracking \cite{Zlatev_WS_99}
properties that allow the scalar field energy density to evolve
alongside the other matter constituents. A problematic feature of
scaling potentials \cite{Ferreira_Joyce_97} is that they do not
lead to accelerative expansion, since the energy density simply
scales with that of matter. Alternatively, certain potentials can
predict a Dark Energy density which alternately dominates the
Universe and decays away; in such models, the acceleration of the
Universe is transient
\cite{Albrecht_Skordis_00,Dodelson_KS_2001,Bento02}. Collectively,
quintessence potentials predict that the density of the Dark
Energy dynamically evolve in time, in contrast to the cosmological
constant. Similar to a cosmological constant, however, the scalar
field is expected to have no significant density perturbations
within the causal horizon, so that they contribute little to the
evolution of the clustering of matter in large-scale structure
\cite{Ferreira_Joyce_98}.

In addition to couplings to ordinary matter, the quintessence
field may have nontrivial couplings to dark matter
\cite{Anderson_Carroll_1997,Farrar_Peebles_2003}. Non perturbative
string-loop effects do not lead to universal couplings, with the
possibility that the dilaton decouples more slowly from dark
matter than it does from gravity and fermions. This coupling can
provide a mechanism to generate acceleration, with a scaling
potential, while also being consistent with Equivalence Principle
tests. It can also explain why acceleration is occurring only
recently, through being triggered by the non-minimal coupling to
the cold dark matter, rather than a feature in the effective
potential \cite{Bean_Magueijo_01, Gasperini_PV_2002}. Such
couplings can not only generate acceleration, but also modify
structure formation through the coupling to CDM density
fluctuations \cite{Bean_01}, in contrast to minimally coupled
quintessence models. Dynamical observables, sensitive to the
evolution in matter perturbations as well as the expansion of the
Universe, such as the matter power spectrum as measured by large
scale surveys, and weak lensing convergence spectra, could
distinguish non-minimal couplings from theories with minimal
effect on clustering. The interaction between dark energy and dark
matter is, of course, present in the generalized Chaplygin gas
model, as in this proposal the fluid has a dual behavior.

It should be noted that for the run-away dilaton scenario
presented in Section \ref{sec:SMT}, comparison with the minimally
coupled scalar field action,

\beq S_{\phi} = {c^3\over 4\pi G}\int d^{4}x\sqrt{-g}
\left[\frac{1}{4}R +\frac{1}{2}\partial_{\mu} \phi\partial^{\mu}
\phi-V(\phi)\right],\eeq

\noindent reveals that the negative scalar kinetic term leads to
an action equivalent to a ``ghost'' in quantum field theory, and
is referred to as ``phantom energy'' in the cosmological context
\cite{Caldwell_02}. Such a scalar field model could in theory
generate acceleration by the field evolving {\it up} the potential
toward the maximum. Phantom fields are plagued by catastrophic UV
instabilities, as particle excitations have a negative mass
\cite{Carroll_HT_03,Cline_JM_04}; the fact that their energy is
unbounded from below allows vacuum decay through the production of
high energy real particles and negative energy ghosts that will be
in contradiction with the constraints on ultra-high energy cosmic
rays \cite{Sreekumar_etal_98}.

Such runaway behavior can potentially be avoided by the
introduction of higher-order kinetic terms in the action.  One
implementation of this idea is ``ghost condensation''
\cite{Arkani_CMZ_04}. Here, the scalar field has a negative
kinetic energy near $\dot\phi=0$, but the quantum instabilities
are stabilized by the addition of higher-order corrections to the
scalar field Lagrangian of the form $(\partial_{\mu}
\phi\partial^{\mu} \phi)^{2}$. The ``ghost'' energy is then
bounded from below, and stable evolution of the dilaton occurs
with $w\ge-1$ \cite{Piazza_Tsujikawa_04}.  The gradient
$\partial_\mu\phi$ is non-vanishing in the vacuum, violating
Lorentz invariance, and may have interesting consequences in
cosmology and in laboratory experiments.

In proposing the scalar field as physical and requiring it to
replace CDM and DE, one has to also calculate how the scalar field
density fluctuations evolve, in order to compare them with density
power spectra from large-scale structure surveys. This is true for
the broader set of phenomenological models including the
generalized Born-Infeld action, associated to the generalized Chaplygin
gas model \cite{Bento2002}. Despite being consistent with
kinematical observations, it has been pointed that they are
disfavored in comparison to the $\Lambda$CDM scenario
\cite{Sandvik_TZ_04,Bean_Dore_03}, even though solutions have been
proposed \cite{Bento04}.

%%%%%%%%%%%%%%%%%%%%%%%%%% Section 5 %%%%%%%%%%%%%%%%%%%%%%%%%%%%%
\section{Gravitational Physics and Experiments in Space}
\label{sec:space}

Recent progress in observational astronomy, astrophysics, and cosmology 
has raised important questions related to gravity and other 
fundamental laws of Nature. There are two approaches 
to physics research in space: one can detect and study signals from remote 
astrophysical objects, 
while the other relies on a carefully designed experiment. 
Although the two methods are 
complementary, the latter has the advantage of utilizing a 
well-understood and controlled laboratory environment in the solar system.
Newly available technologies in conjunction with existing space 
capabilities offer unique 
opportunities to take full advantage of the variable gravity potentials, 
large heliocentric 
distances, and high velocity and acceleration regimes that are present 
in the solar system.  
As a result, solar system experiments can significantly advance our 
knowledge of fundamental 
physics and are capable of providing the missing links connecting 
quarks to the cosmos.

In this section we will discuss theoretical motivation of and 
innovative ideas for the 
advanced gravitational space experiments.

\subsection{Testable Implications of Recent Theoretical Proposals}

The theories that were discussed in the previous section offer a
diverse range of characteristic experimental predictions differing
from those of GR that would allow their falsification. The most
obvious tests would come from the comparison of the predictions of
the theory to astrophysical and cosmological observations where
the dynamics are dominated by very small gravitational fields. As
a result, one might expect that these mechanisms would lead to
small effects in the motion of the bodies in the solar system,
short- and long-scale modifications of Newton's law, as well as
astrophysical phenomena.

Below we will discuss these possible tests and estimate the sizes of the 
expected effects.

\subsubsection{Testing Newton's Law at Short Distances}

It was observed that many recent theories predict observable
experimental signatures in experiments testing Newton's law at
short distances. For instance in the case of MOND-inspired
theories discussed in Section~\ref{sec:dm-grav-mod}, there may be
an extra scalar excitation of the spacetime metric besides the
massless graviton. Thus, in the effective gravitational theory
applicable to the terrestrial conditions, besides the massless
spin two graviton, one would also have an extra scalar field with
gravitational couplings and with a small mass. A peculiar feature
of such a local effective theory on a Schwarzschild background is
that there will be a preferred direction that will be reflected in
an anisotropy of the force that this scalar excitation will
mediate. For an experiment conducted in the terrestrial conditions
one expects short ranges modifications of Newton's law at
distances of $ \sim 0.1$ mm, regime that is close to that already
being explored in some laboratory experiments
\cite{Adelberger_etal_2003a,Adelberger_etal_2003b}.

For an experiment on an Earth-orbiting platform, one explores
another interesting regime for which the solar mass and the
Sun-Earth distance are the dominant factors in estimating the size
of the effects. In this case the range of interest is $\sim 10^4$~m. 
However in measuring the gravitational field of an object one
has to measure this field at a distance that is larger than the
critical distance for which the self-shielding of the extra scalar
excitation induced by the object itself is enough to switch off
the modification. This means that, for an experiment in the inner
solar system, we could only see significant modifications in the
gravitational field of objects whose characteristic distance is
smaller than $10^4$ m, thus limiting the mass of the body to be
below $\sim 10^{9}$ kg. As an example, one can place an object
with mass of $10^3$ kg placed on a heliocentric orbit at $\sim 1$~AU distance. 
For this situation, one may expect modifications of
the body's gravitational field at distances within the range of
$\sim 10-10^4$ m.  Note that at shorter distances the scalar
effectively decouples because of the self-gravitational effect of
the test object; also, at longer distances the mass induced by
solar gravitational field effectively decouples the scalar.

\subsubsection{Solar System Tests of Relativistic Gravity}

Although many effects expected by gravity modification models are
suppressed within the solar system, there are measurable effects
induced by some long-distance modifications of gravity (notably
the DGP model \cite{Dvali_GP_2000})). For instance, in the case of
the precession of the planetary perihelion in the solar system,
the anomalous perihelion advance, $\Delta \phi$, induced by a
small correction, $\delta V_N$, to Newton's potential, $V_N$, is
given in radians per revolution \cite{Dvali:2002vf}  by
\beq \Delta \phi \simeq \pi r
{d\over dr}\left(r^2 {d \over dr}\left({\delta
V_N \over rV_N}\right)\right). \eeq

The most reliable data regarding the planetary perihelion advances come
from the inner planets of the solar system \cite{Pitjeva}, where
most of the corrections are negligible. However, with its
excellent 2-cm-level range accuracy \cite{Williams_Turyshev_Boggs_2004}, LLR offers an interesting possibility to test for these new effects. Evaluating the expected magnitude of the effect to the Earth-Moon system, one predicts an anomalous shift of $\Delta \phi \sim 10^{-12}$, to be compared with the achieved accuracy of $2.4\times 10^{-11}$ \cite{Dvali:2002vf}. Therefore, the theories of gravity modification result in an intriguing possibility of discovering new physics, if one focuses on achieving higher precision in modern astrometrical measurements; this accuracy increase is within the reach and should be attempted in the near future.

The quintessence models discussed in Section
\ref{sec:sc-models-de} offer the possibility of observable
couplings to ordinary matter, makes these models especially
attractive for the tests even on the scales of the solar system.
Even if we restrict attention to non-renormalizable couplings
suppressed by the Planck scale, tests from fifth-force experiments
and time-dependence of the fine-structure constant imply that such
interactions must be several orders of magnitude less than
expected \cite{Carroll_1998}.  Further improvement of existing
limits on violations of the Equivalence Principle in terrestrial
experiments and also in space would also provide important constraints on
dark-energy models.

Another interesting experimental possibility is provided by the
``chameleon'' effect \cite{Khoury_Weltman_04,Brax_etal_04}. Thus,
by coupling to the baryon energy density, the scalar field value
can vary across space from solar system to cosmological scales.
Though the small variation of the coupling on Earth satisfies the
existing terrestrial experimental bounds, future gravitational
experiments in space such as measurements of variations in the
gravitational constant or test of Equivalence Principle, may
provide critical information for the theory.

There is also a possibility that the dynamics of the quintessence
field evolves to a point of minimal coupling to matter. In
\cite{Damour_Polyakov_1994a} it was shown that  $\phi$ could be
attracted towards a value $\phi_{m}(x)$ during the matter
dominated era that decoupled the dilaton from matter. For
universal coupling, $f(\varphi)=g(\varphi)=q_{i}(\varphi)$ (see
Eq.~\ref{eq:sc-tensor}), this would motivate for improving the
accuracy of the equivalence principle and other tests of GR.
Ref.~\cite{Veneziano_2001} suggested that with a large number of
non-self-interacting matter species, the coupling constants are
determined by the quantum corrections of the matter species, and
$\phi$ would evolve as a run-away dilaton with asymptotic value
$\phi_{m}\rightarrow\infty$. More recently, in Refs.
\cite{DPV02a} the quantity $\frac{1}{2}(1-\gamma)$ has been
estimated, within the framework compatible with string theory and
modern cosmology, which basically confirms the previous result
\cite{Damour_Nordtvedt_1993a}. This recent
analysis discusses a scenario where a composition-independent
coupling of dilaton to hadronic matter produces detectable
deviations from GR in high-accuracy light deflection experiments
in the solar system. This work assumes only some general property
of the coupling functions and then only assume that $(1-\gamma)$
is of order of one at the beginning of the controllably classical
part of inflation. It was shown  in \cite{DPV02a} that one can
relate the present value of $\frac{1}{2}(1-\gamma)$ to the
cosmological density fluctuations; the level of the expected
deviations from GR is $\sim0.5\times10^{-7}$ \cite{DPV02a}. Note
that these predictions are based on the work in scalar-tensor
extensions of gravity which are consistent with, and indeed often
part of, present cosmological models provide a strong motivation
for improvement of the accuracy of gravitational tests in the
solar system.

\subsubsection{Observations on Astrophysical and Cosmological Scales}

The new theories also suggest an interesting observable effects on
astrophysical and cosmological observations (see for instance
\cite{Aguirre:2003pg}). In this respect, one can make unambiguous
predictions for the rotation curves of spiral galaxies with the
mass-to-light ratio being the only free parameter. Specifically,
it has been argued that a skew-symmetric field with a suitable
potential could account for galaxy and cluster rotation curves
\cite{moffat05}. One can even choose an appropriate potential that
would then give rise to flat rotation curves that obey the
Tully-Fisher law \cite{Tully:1977fu}. But also other aspects of
the observations of galactic dynamics can be used to constrain a
MOND-like modification of Newton's potential (see
\cite{Zhao:2005zq}). And notice also that our theory violates the
strong equivalence principle, as expected for any relativistic
theory for MOND \cite{Milgrom:1983ca}, since locally physics will
intrinsically depend on the background gravitational field. This
will be the case if the background curvature dominates the
curvature induced by the local system, similarly to the ``external
field effect'' in MOND.

At larger scales, where one can use the equivalence with a
scalar-tensor theory more reliably, one can then compare the
theory against the observations of gravitational lensing in
clusters, the growth of large scale structure and the fluctuations
of the CMBR. In fact, it has been pointed out that if GR was
modified at large distances, an inconsistency between the allowed
regions of parameter space would allow for Dark Energy models
verification when comparing the bounds on these parameters
obtained from CMBR, and large scale structure \cite{Ishak:2005zs}.
This means that although some cosmological observables, like the
expansion history of the Universe, can be indistinguishable in
modified gravity and Dark Energy models, this degeneracy is broken
when considering other cosmological observations and in particular
the growth of large scale structure and the Integrated Sachs-Wolfe
effect (ISW) have been shown to be good discriminators for models
in which GR is modified \cite{Zhang:2005vt}. It has been recently
pointed out that the fact that in the DGP model the effective
Newton's constant increases at late times as the background
curvature diminishes, causes a suppression of the ISW that brings
the theory into better agreement with the CMBR data than the
$\Lambda$CDM model \cite{Sawicki:2005cc}.

\subsection{New Experiments and Missions}

Theoretical motivations presented above have stimulated development of 
several highly-accurate space experiments. Below we will briefly discuss 
science objectives and experimental design for several advanced experiments, 
namely MICROSCOPE, STEP, and HYPER missions, APOLLO LLR facility, and the LATOR mission.

\subsubsection{MICROSCOPE, STEP, and HYPER Missions}
\label{missions}

Ground experiments designed to verify the validity of the WEP are
limited by unavoidable microseismic activity of Earth, while the
stability of space experiments offers an improvement in the
precision of current tests by a factor of $10^6$. Most probably,
the first test of the WEP in space will be carried out by the
MICROSCOPE (MICROSatellite a traine Compensee pour l'Observation
du Principe d'Equivalence) mission led by CNES and ESA. The
drag-free MICROSCOPE satellite, transporting two pairs of test
masses, will be launched into a sun-synchronous orbit at
$600$~km altitude. The differential displacements between each
test masses will of a pair be measured by capacitive sensors at
room-temperature, with an expected precision of one part in
$10^{15}$.

The more ambitious joint ESA/NASA STEP (Satellite Test of the
Equivalence Principle) mission which is proposed to be launched
in the near future into a circular, sun-synchronous orbit with altitude of $600~km$. The drag-free STEP spacecraft will carry four
pairs of test masses stored in a dewar of superfluid He at a 2~K
temperature. Differential displacements between the test masses of
a pair will be measured by SQUID (Superconducting QUantum
Interference Device) sensors, testing the WEP with an expected
precision of $\Delta a/a \sim 10^{18}$.

Another quite interesting test of the WEP involves atomic
interferometry: high-precision gravimetric measurements can be
taken via the interferometry of free-falling caesium atoms, and
such a concept has already yielded a precision of 7 parts per
$10^9$ \cite{Peters}. This can only be dramatically improved in
space, through a mission like HYPER (HYPER-precision cold atom
interferometry in space). ESA's HYPER spacecraft would be in a
sun-synchronous circular orbit at $700$~km altitude. Two atomic
Sagnac units are to be accommodated in the spacecraft, comprising
four cold atom interferometers able to measure rotations and
accelerations along two orthogonal planes. By comparing the rates
of fall of caesium and rubidium atoms, the resolution of the atom
interferometers of the HYPER experiment could, in principle, test
the WEP with a precision of one part in $10^{15}$ or $10^{16}$
\cite{HYPER}.

It is worth mentioning that proposals have been advanced to test
the WEP by comparing the rate of fall of protons and antiprotons
in a cryogenic vacuum facility that will be available at the ISS
\cite{Lewis}. The concept behind this Weak Equivalence Antimatter
eXperiment (WEAX) consists of confining antiprotons for a few
weeks in a Penning trap, in a geometry such that gravity would
produce a perturbation on the motion of the antiprotons. The
expected precision of the experiment is of one part in $10^6$,
three orders of magnitude better than for a ground experiment.

It is clear that testing the WEP in space requires pushing current
technology to the limit; even though no significant violations of
this principle are expected, any anomaly would provide significant
insight into new and fundamental physical theories. The broad
perspectives and the potential impact of testing fundamental
physics in space were discussed in Ref. \cite{Bertolami20}.

\subsubsection{APOLLO -- a $mm$-class LLR Facility}

The Apache Point Observatory Lunar Laser-ranging Operation is a
new LLR effort designed to achieve millimeter range precision and
corresponding order-of-magnitude gains in measurements of
fundamental physics parameters. The APOLLO project design and
leadership responsibilities are shared between the University of
California at San Diego and the University of Washington. In
addition to the modeling aspects related to this new LLR
facility, a brief description of APOLLO and associated
expectations is provided here for reference. A more complete
description can be found in \cite{Murphy_etal_2002}.

The principal technologies implemented by APOLLO include a robust
Nd:YAG laser with 100~ps pulse width, a GPS-slaved 50 MHz
frequency standard and clock, a 25~ps-resolution time interval
counter, and an integrated avalanche photo-diode (APD) array. The
APD array, developed at Lincoln Labs, is a new technology that
will allow multiple simultaneous photons to be individually
time-tagged, and provide two-dimensional spatial information for
real-time acquisition and tracking capabilities.

The overwhelming advantage APOLLO has over current LLR operations
is a 3.5 m astronomical quality telescope at a good site. The site
in the Sacramento Mountains of southern New Mexico offers high
altitude (2780~m) and very good atmospheric ``seeing'' and image
quality, with a median image resolution of $1.1$ arcseconds. Both
the image sharpness and large aperture enable the APOLLO
instrument to deliver more photons onto the lunar retroreflector
and receive more of the photons returning from the reflectors,
respectively. Compared to current operations that receive, on
average, fewer than 0.01 photons per pulse, APOLLO should be well
into the multi-photon regime, with perhaps 5-10 return photons per
pulse. With this signal rate, APOLLO will be efficient at finding
and tracking the lunar return, yielding hundreds of times more
photons in an observation than current operations deliver. In
addition to the significant reduction in statistical error
($\sim\sqrt{N}$ reduction), the high signal rate will allow
assessment and elimination of systematic errors in a way not
currently possible.

The new LLR capabilities offered by the newly developed APOLLO
instrument offer a unique opportunity to improve accuracy of a
number of fundamental physics tests. The APOLLO project will push
LLR into the regime of millimetric range precision which
translates to an order-of-magnitude improvement in the
determination of fundamental physics parameters. For the Earth and
Moon orbiting the Sun, the scale of relativistic effects is set by
the ratio $(GM / r c^2)\sim v^2 /c^2 \sim 10^{-8}$.  Relativistic
effects are small compared to Newtonian effects.  The APOLLO's
1 mm range accuracy corresponds to $3\times 10^{-12}$ of the
Earth-Moon distance.  The resulting LLR tests of gravitational
physics would improve by an order of magnitude: the Equivalence
Principle would give uncertainties approaching $10^{-14}$, tests
of GR effects would be $<0.1$\%, and estimates of
the relative change in the gravitational constant would be 0.1\%
of the inverse age of the Universe. This last number is impressive
considering that the expansion rate of the Universe is
approximately one part in 10$^{10}$ per year.

Therefore, the gain in our ability to conduct even more precise
tests of fundamental physics is enormous, thus this new instrument
stimulates development of better and more accurate models for the
LLR data analysis at a mm-level \cite{llr-ijmpd}.  

\subsubsection{The LATOR Mission}

The recently proposed Laser Astrometric Test Of Relativity (LATOR)
\cite{Laser_Clocks_LATOR,solvang_lator04,Lator01,Texas@Stanford_lator}
is an experiment designed to test the metric nature of gravitation
-- a fundamental postulate of Einstein's theory of general
relativity. By using a combination of independent time-series of
highly accurate gravitational deflection of light in the immediate
proximity to the sun, along with measurements of the Shapiro time
delay on interplanetary scales (to a precision respectively better
than $10^{-13}$ radians and 1 cm), LATOR will significantly
improve our knowledge of relativistic gravity. The primary mission
objective is to i) measure the key post-Newtonian Eddington
parameter $\gamma$ with accuracy of a part in 10$^9$. The quantity
$(1-\gamma)$ is a direct measure for presence of a new interaction
in gravitational theory, and, in its search, LATOR goes a factor
$30,000$ beyond the present best result, Cassini's 2003 test. Other
mission objectives include: ii) first measurement of gravity's
non-linear effects on light to $\sim 0.01\%$ accuracy; including
both the traditional Eddington $\beta$ parameter via gravity
effect on light to $\sim0.01\%$ accuracy and also the spatial
metric's 2-nd order potential contribution $\delta$ (never measured
before); iii) direct measurement of the solar quadrupole moment
$J_2$ (currently unavailable) to accuracy of a part in 200 of its
expected size; iv) direct measurement of the ``frame-dragging''
effect on light due to the sun's rotational gravitomagnetic field,
to $0.1\%$ accuracy. LATOR's primary measurement pushes to
unprecedented accuracy the search for cosmologically relevant
scalar-tensor theories of gravity by looking for a remnant scalar
field in today's solar system. The key element of LATOR is a
geometric redundancy provided by the laser ranging and
long-baseline optical interferometry.

As a result, LATOR will be able to test the metric nature of the
Einstein's general theory of relativity in the most intense
gravitational environment available in the solar system -- the
extreme proximity to the sun. It will also test alternative
theories of gravity and cosmology, notably scalar-tensor theories,
by searching for cosmological remnants of scalar field in the
solar system. LATOR will lead to very robust advances in the tests
of fundamental physics: this mission could discover a violation or
extension of GR, or reveal the presence of an
additional long range interaction in the physical law. There are
no analogs to the LATOR experiment; it is unique and is a natural
culmination of solar system gravity experiments \cite{Laser_Clocks_LATOR}.

LATOR mission is the 21st century version of Michelson-Morley-type
experiment searching for a cosmologically evolved scalar field in
the solar system. In spite of the previous space missions
exploiting radio waves for tracking the spacecraft, this mission
manifests an actual breakthrough in the relativistic gravity
experiments as it allows to take full advantage of the optical
techniques that recently became available. LATOR has a number of
advantages over techniques that use radio waves to measure
gravitational light deflection. Thus, optical technologies allow
low bandwidth telecommunications with the LATOR spacecraft. The
use of the monochromatic light enables the observation of the
spacecraft at the limb of the sun. The use of narrowband filters,
coronagraph optics and heterodyne detection will suppress
background light to a level where the solar background is no
longer the dominant noise source. The short wavelength allows much
more efficient links with smaller apertures, thereby eliminating
the need for a deployable antenna. Finally, the use of the ISS
enables the test above the Earth's atmosphere -- the major source
of astrometric noise for any ground based interferometer. This
fact justifies LATOR as a space mission. LATOR is envisaged as a
partnership between European and US institutions and with clear
areas of responsibility between the space agencies: NASA provides
the deep space mission components, while optical infrastructure on
the ISS would be an ESA contribution.

%%%%%%%%%%%%%%%%%%%%%%%%%% Section 6 %%%%%%%%%%%%%%%%%%%%%%%%%%%%%
\section*{Conclusions}
\label{sec:conclusions}

General theory of relativity is one of the most elegant theories of physics; 
it is also one of the most empirically verified. Thus, almost ninety years of 
testing have also proved that GR has so far successfully accounted for 
all encountered phenomena and experiments in the solar system and with binary pulsars.
However, despite that there are predictions of the theory that require still 
confirmation and detailed analysis, most notably the direct detection of gravitational waves.
However, there are new  motivations to test the theory to even a higher precisions 
that already led to a number of experimental proposals to advance the knowledge of 
fundamental laws of physics.

Recent progress in observational astronomy, astrophysics, and cosmology has raised important questions 
related to gravity and other fundamental laws of Nature.  There are two approaches to physics 
research in space: one can detect and study signals from remote astrophysical objects, while 
the other relies on a carefully designed experiment.  Although the two methods are 
complementary, the latter has the advantage of utilizing a well-understood and controlled 
laboratory environment in the solar system.

Newly available technologies in conjunction with existing space capabilities offer unique 
opportunities to take full advantage of the variable gravity potentials, large heliocentric 
distances, and high velocity and acceleration regimes that are present in the solar system.
A common feature of precision gravity experiments is that they must operate in the noise 
free environment needed to achieve the ever increasing accuracy. These requirements are 
supported by the progress in the technologies, critical for space  exploration, namely 
the highly-stable, high-powered, and space-qualified lasers, highly-accurate frequency 
standards, and the drag-free technologies.
This progress advances both the science and technology for the laboratory experiments 
in space with laboratory being the entire solar system. As a result, solar system 
experiments can significantly advance our knowledge of fundamental physics and are 
capable of providing the missing links connecting quarks to the cosmos.

Concluding, it is our hope that the recent progress will lead to establishing a more 
encompassing theory to describe all physical interactions in an unified fashion that 
harmonizes the spacetime description of GR with quantum mechanics. 
This unified theory is needed to address many of the standing difficulties we face 
in theoretical physics: Are singularities an unavoidable property of spacetime? What is 
the origin of our Universe? How to circumvent the cosmological constant problem and 
achieve a successful period of inflation and save our Universe from an embarrassing 
set of initial conditions? The answer to these questions is, of course, closely 
related to the nature of gravity. It is an exciting prospect to think that experiments 
carried out in space will be the first to provide the essential insights on the brave 
new world of the new theories to come.

%**************************************

\subsubsection*{Acknowledgments~~}
The work of SGT described was carried out at the Jet Propulsion
Laboratory, California Institute of Technology, under a contract
with the National Aeronautics and Space Administration.

\bibliographystyle{unstr}

\end{document}